\documentclass[aoas,MSNbibl,nameyear,dvips]{arximspdf}
\usepackage{algorithm,algorithmic}
\usepackage{eqparbox}
\usepackage{graphicx}

\doi{10.1214/14-AOAS742} 
\volume{8}
\issue{3}
\pubyear{2014}
\firstpage{1281}
\lastpage{1313}
\docsubty{FLA}

\makeatletter

\renewcommand{\mid}{|}
\newcommand{\hatBFsub}[2]{\hat{\bolds{#1}}_{#2}}
\newcommand{\symsubsupB}[3]{{\mathbf{#1}}_{#2}^{(#3)}}
\newcommand{\symsubsupT}[3]{{\mathbf{#1}}_{#2}^{(#3)^T}}
\makeatother

\begin{document}
\begin{frontmatter}

\title{Joint modeling of multiple time series via the beta process
with application to motion capture~segmentation}
\runtitle{Joint modeling of multiple time series via the beta process}

\begin{aug}
\author[A]{\fnms{Emily B.} \snm{Fox}\corref{}\thanksref{T1,WASH}\ead[label=e1]{ebfox@stat.washington.edu}},
\author[B]{\fnms{Michael C.} \snm{Hughes}\thanksref{T1,T2,BROWN}\ead[label=e2]{mhughes@cs.brown.edu}},\\
\author[B]{\fnms{Erik B.} \snm{Sudderth}\thanksref{T1,BROWN}\ead[label=e3]{sudderth@cs.brown.edu}}
\and
\author[C]{\fnms{Michael I.} \snm{Jordan}\thanksref{T1,BERK}\ead[label=e4]{jordan@stat.berkeley.edu}}
\runauthor{Fox, Hughes, Sudderth and Jordan}
\affiliation{University of Washington\thanksmark{WASH}, Brown
University\thanksmark{BROWN}\\ and University of California, Berkeley\thanksmark{BERK}}
\address[A]{E. B. Fox\\
Department of Statistics\\
University of Washington\\
Box 354322\\
Seattle, Washington 98195\\
USA\\
\printead{e1}}
\address[B]{M. C. Hughes\\
E. B. Sudderth\\
Department of Computer Science\\
Brown University\\
115 Waterman St., Box 1910\\
Providence, Rhode Island 02912\\
USA\\
\printead{e2}\\
\phantom{E-mail: }\printead*{e3}}
\address[C]{M. I. Jordan\\
Department of Statistics\\
\quad and Department of EECS\\
University of California, Berkeley\\
427 Evans Hall\\
Berkeley, California 94720\\
USA\\
\printead{e4}}
\end{aug}
\thankstext{T1}{Supported in part by AFOSR Grant FA9550-12-1-0453 and
ONR Contracts/Grants N00014-11-1-0688 and N00014-10-1-0746.}
\thankstext{T2}{Supported in part by an NSF Graduate Research
Fellowship under Grant DGE0228243.}

\received{\smonth{5} \syear{2013}}
\revised{\smonth{1} \syear{2014}}

%
\begin{abstract}
We propose a Bayesian nonparametric approach to the problem of jointly
modeling multiple related time series. Our model discovers a latent set
of dynamical behaviors shared among the sequences, and segments each
time series into regions defined by a subset of these behaviors. Using
a beta process prior, the size of the behavior set and the sharing
pattern are both inferred from data. We develop Markov chain Monte
Carlo (MCMC) methods based on the Indian buffet process representation
of the predictive distribution of the beta process. Our MCMC inference
algorithm efficiently adds and removes behaviors via novel split-merge
moves as well as data-driven birth and death proposals, avoiding the
need to consider a truncated model. We demonstrate promising results on
unsupervised segmentation of human motion capture data.
\end{abstract}

%
\end{frontmatter}

\setcounter{footnote}{2}
\section{Introduction}
\begin{figure}

\includegraphics{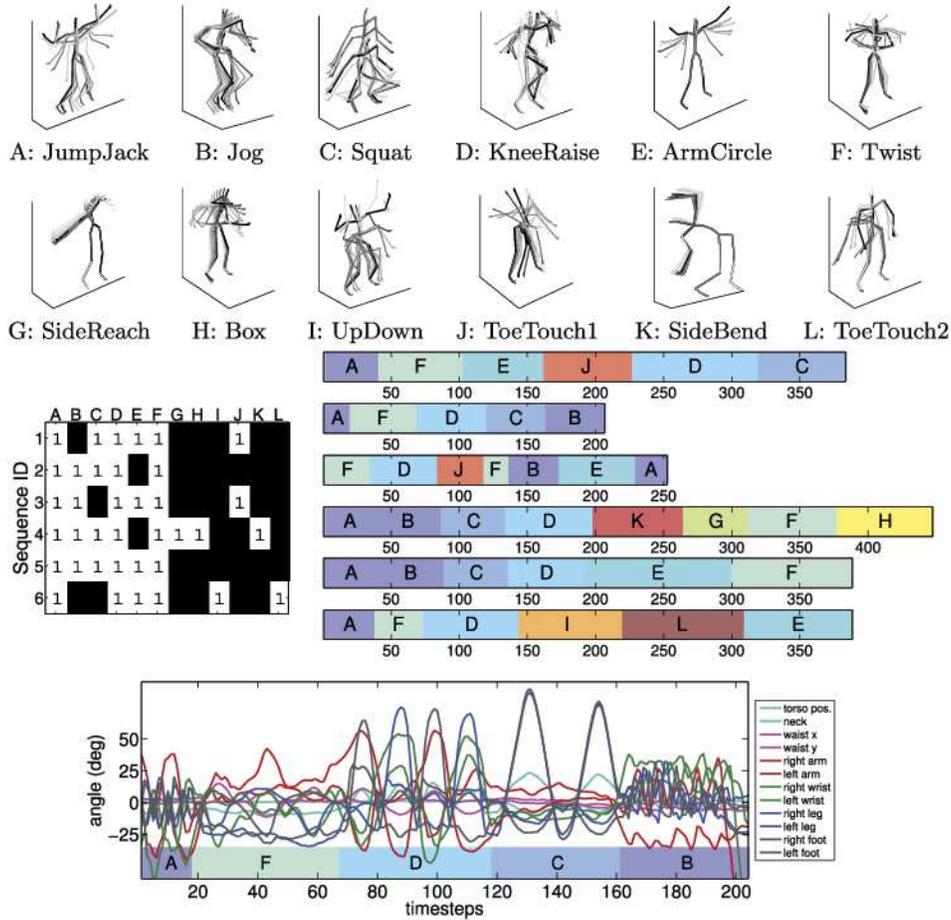}

\caption{Motivating data set:
6 sequences of motion capture data [\citet{CMUmocap}], with manual annotations.
\emph{Top}: Skeleton visualizations of 12 possible exercise behavior types observed across all sequences.
\emph{Middle left}: Binary feature assignment matrix $\mathbf{F}$ produced by manual annotation. Each row indicates which exercises are present in a particular sequence.
\emph{Middle right}: Discrete segmentations $\mathbf{z}$ of all six time series into the 12 possible exercises, produced by manual annotation.
\emph{Bottom}: Sequence 2's observed multivariate time series data. Motion capture sensors measure 12 joint angles every 0.1 seconds.
\emph{Proposed model}: The BP-AR-HMM takes as input the observed time series sensor data across multiple sequences. It aims to recover the global behavior set, the binary assignments $\mathbf{F}$, and the detailed segmentations $\mathbf{z}$. When segmenting each sequence, our model
only uses behaviors which are present in the corresponding row of~$\mathbf{F}$.}\label{figGroundTruthMocap6}
\end{figure}

Classical time series analysis has generally focused on the study of a
single (potentially multivariate) time series. Instead, we consider
analyzing \emph{collections} of related time series, motivated by the
increasing abundance of such data in many domains. In this work we
explore this problem by considering time series produced by motion
capture sensors on the joints of people performing exercise routines.
An individual recording provides a multivariate time series that can be
segmented into types of exercises (e.g., jumping jacks, arm-circles,
and twists). Each exercise type describes locally coherent and simple
dynamics that persist over a segment of time. We have such motion
capture recordings from \emph{multiple} individuals, each of whom
performs some subset of a global set of exercises, as shown in
Figure~\ref{figGroundTruthMocap6}. Our goal is to discover the set of
global exercise types (``behaviors'') and their occurrences in each
individual's data stream. We would like to take advantage of the
overlap between individuals: if a jumping-jack behavior is discovered
in one sequence, then it can be used to model data for other
individuals. This allows a combinatorial form of shrinkage involving
subsets of behaviors from a global collection.

%

A flexible yet simple method of describing single time series with such
patterned behaviors is the class of \emph{Markov switching processes}.
These processes assume that the time series can be described via Markov
transitions between a set of latent dynamic behaviors which are
individually modeled via temporally independent linear dynamical
systems. Examples include the hidden Markov model (HMM), switching
vector autoregressive (VAR) process, and switching linear dynamical
system (SLDS). These models have proven useful in such diverse fields
as speech recognition, econometrics, neuroscience, remote target
tracking, and human motion capture. In this paper, we focus our
attention on the descriptive yet computationally tractable class of
switching VAR processes. Here, the state of the underlying Markov
process encodes the behavior exhibited at a given time step, and each
dynamic behavior defines a VAR process. That is, conditioned on the
Markov-evolving state, the likelihood is simply a VAR model with
time-varying parameters.

To discover the dynamic behaviors shared between multiple time series,
we propose a feature-based model. The entire collection of time series
can be described by a globally shared set of possible behaviors.
Individually, however, each time series will only exhibit a subset of
these behaviors. The goal of joint analysis is to discover which
behaviors are shared among the time series and which are unique. We
represent the behaviors possessed by time series $i$ with a binary
\emph{feature vector}~$\mathbf{f}_{ i}$, with $f_{ik}=1$ indicating
that time series $i$ uses global behavior $k$ (see Figure~\ref
{figGroundTruthMocap6}). We seek a prior for these feature vectors
which allows flexibility in the number of behaviors and encourages the
sharing of behaviors. Our desiderata motivate a feature-based Bayesian
nonparametric approach based on the \emph{beta process}~[\citet
{Hjort90,Thibaux07}]. Such an approach allows for \emph{infinitely}
many potential behaviors, but encourages a sparse representation. Given
a fixed feature set, our model reduces to a collection of finite
Bayesian VAR processes with partially shared parameters.
%

We refer to our model as the \emph{beta-process autoregressive hidden
Markov model}, or BP-AR-HMM. We also consider a simplified version of
this model, referred to as the BP-HMM, in which the AR emission models
are replaced with a set of conditionally independent emissions.
Preliminary versions of these models were partially described in
\citet{BPHMM} and in \citet{BPHMMNIPS12}, who developed
improved Markov chain Monte Carlo (MCMC) inference procedures for the
BP-AR-HMM. In the current article we provide a unified and
\mbox{comprehensive} description of the model and we also take further steps
toward the development of an efficient inference algorithm for the
BP-AR-HMM. In particular, the \emph{unbounded} nature of the set of
possible behaviors available to our approach presents critical
challenges during posterior inference. To efficiently explore the
space, we introduce two novel MCMC proposal moves: (1) split-merge
moves to efficiently change the feature structure for many sequences at
once, and (2) data-driven reversible jump moves to add or delete
features unique to one sequence. We expect the foundational ideas
underlying both contributions (split-merge and data-driven birth--death)
to generalize to other nonparametric models beyond the time-series
domain. Building on an earlier version of these ideas in \citet
{BPHMMNIPS12}, we show how to perform data-driven birth--death proposals
using only discrete assignment variables (marginalizing away continuous
HMM parameters), and demonstrate that annealing the Hastings term in
the acceptance ratio can dramatically improve performance.

Our presentation is organized as follows. Section~\ref{secdata}
introduces motion capture data. In Section~\ref{secmodel} we present
our proposed beta-process-based model for multiple time series.
Section~\ref{secprior} provides a formal specification of all prior
distributions, while Section~\ref{secoverview} summarizes the model.
Efficient posterior computations based on an MCMC algorithm are
developed in Section~\ref{secMCMC}. The algorithm does not rely on
model truncation; instead, we exploit the finite dynamical system
induced by a fixed set of features to sample efficiently, while using
data-driven reversible jump proposals to explore new features.
Section~\ref{secSplitMerge} introduces our novel split-merge
proposals, which allow the sampler to make large-scale improvements
across many variables simultaneously. In Section~\ref{secrelated} we
describe related work. Finally, in Section~\ref{secMoCap} we present
results on unsupervised segmentation of data from the CMU motion
capture database~[\citet{CMUmocap}]. Further details on our
algorithms and experiments are available in the supplemental
article~[\citet{BPARHMMSupp}].


\section{Motion capture data}\label{secdata}
Our data consists of motion capture recordings taken from the CMU MoCap
database (\url{http://mocap.cs.cmu.edu}). From the available set of 62
positions and joint angles, we examine 12 measurements deemed most
informative for the gross motor behaviors we wish to capture: one body
torso position, one neck angle, two waist angles, and a symmetric pair
of right and left angles at each subject's shoulders, wrists, knees,
and feet. As such, each recording provides us with a 12-dimensional
time series. A collection of several recordings serves as the observed
data which our model analyzes.

An example data set of six sequences is shown in Figure~\ref
{figGroundTruthMocap6}. This data set contains three sequences from
Subject 13 and three from Subject 14. These sequences were chosen
because they had many exercises in common, such as ``squat'' and
``jog,'' while also containing several unique behaviors appearing in
only one sequence, such as ``side bend.'' Additionally, we have human
annotations of these sequences, identifying which of 12 exercise
behaviors was present at each time step, as shown in Figure~\ref
{figGroundTruthMocap6}. These human segmentations serve as
ground-truth for assessing the accuracy of our model's estimated
segmentations (see Section~\ref{secMoCap}). In addition to analyzing
this small data set, we also consider a much larger 124 sequence data
set in Section~\ref{secMoCap}.

\section{A featural model for relating multiple time series} \label{secmodel}
In our applications of interest, we are faced with a \emph{collection}
of $N$ time series representing realizations of related dynamical
phenomena. Our goal is to discover dynamic behaviors shared between the
time series. Through this process, we can infer how the data streams
relate to one another as well as harness the shared structure to pool
observations from the same behavior, thereby improving our estimates of
the dynamic parameters.

We begin by describing a model for the dynamics of each individual time
series. We then describe a mechanism for representing dynamics which
are shared between multiple data streams. Our Bayesian nonparametric
prior specification plays a key role in this model, by addressing the
challenge of allowing for uncertainty in the number of dynamic
behaviors exhibited within and shared across data streams.
%
\subsection{Per-series dynamics}
\label{secMarkovSwitchingProcesses}
We model the dynamics of each time series as a \emph{Markov switching
process} (MSP). Most simply, one could consider a hidden Markov model
(HMM)~[\citet{Rabiner89}]. For observations $\mathbf{y}_t \in
\mathbb{R}^d$ and hidden state $z_t$, the HMM assumes
%
%
\begin{eqnarray}\label{eqnHMMmode}
z_t\mid z_{t-1} &\sim&\pi_{z_{t-1}},
\nonumber
\\[-8pt]
\\[-8pt]
\mathbf{y}_t \mid z_t &\sim& F(\theta_{z_t}),
\nonumber
\end{eqnarray}
for an indexed family of distributions $F(\cdot)$. Here, $\pi_k$ is
the state-specific \emph{transition distribution} and $\theta_k$ the
\emph{emission parameters} for state~$k$.

The modeling assumption of the HMM that observations are conditionally
independent given the latent state sequence is insufficient to capture
the temporal dependencies present in human motion data streams.
Instead, one can assume that the observations have \emph{conditionally
linear dynamics}. Each latent HMM state then models a single linear
dynamical system, and over time the model can switch between dynamical
modes by switching among the states. We restrict our attention in this
paper to switching vector autoregressive (VAR) processes, or \emph
{autoregressive HMMs} (AR-HMMs), which are both broadly applicable and
computationally practical.

We consider an AR-HMM where, conditioned on the latent state $z_t$, the
observations evolve according to a state-specific order-$r$ VAR
process:\footnote{We denote an order-$r$ VAR process by VAR($r$).}
%
%
\begin{equation}
\mathbf{y}_t = \sum_{\ell=1}^r
A_{\ell,z_t}\mathbf{y}_{t-\ell} + \mathbf{e}_t(z_t)
= \mathbf{A}_k \tilde{\mathbf{y}}_{t} + \mathbf
{e}_t(z_t), \label{eqnSVAR}
\end{equation}
where $\mathbf{e}_t(z_t) \sim\mathcal{N}(0,\Sigma_{z_t})$ and
$\tilde{\mathbf{y}}_{t} =
[\matrix{
\smash{\mathbf{y}_{t-1}^T} & \cdots& \smash{\mathbf{y}_{t-1}^T}
}]^T$ are the aggregated past observations. We refer to $\mathbf{A}_k =
[\matrix{
A_{1,k} & \cdots& A_{r,k}
}]
$ as the set of \emph{lag matrices}. Note that an HMM with zero-mean
Gaussian emissions arises as a special case of this model when $\mathbf
{A}_{k}=\mathbf{0}$ for all~$k$. Throughout, we denote the VAR parameters
for the $k$th state as $\theta_k = \{\mathbf{A}_k,\Sigma_k\}$ and refer
to each VAR process as a \emph{dynamic behavior}. For example, these
parameters might each define a linear motion model for the behaviors
\emph{walking}, \emph{running}, \emph{jumping}, and so on; our time
series are then each modeled as Markov switches between these
behaviors. We will sometimes refer to $k$ itself as a ``behavior,''
where the intended meaning is the VAR model parameterized by $\theta_k$.
%
\subsection{Relating multiple time series}\label{secmultipleTimeSeries}
There are many ways in which a collection of data streams may be \emph
{related}. In our applications of interest, our $N$ time series are
related by the overlap in the set of dynamic behaviors each exhibits.
Given exercise routines from $N$ actors, we expect both sharing and
variability: some people may switch between walking and running, while
others switch between running and jumping. Formally, we define a \emph
{shared} set of dynamic behaviors $\{\theta_1,\theta_2,\ldots\}$. We
then associate some subset of these behaviors with each time series $i$
via a binary \emph{feature} vector \mbox{$\mathbf{f}_{ i} = [f_{i1},
f_{i2}, \ldots]$}. Setting $f_{ik}=1$ implies that time series $i$
exhibits behavior~$k$ for some subset of values $t \in\{1,\ldots,T_i\}
$, where $T_i$ is the length of the $i$th time series.

%
%
\begin{figure}[b]

\includegraphics{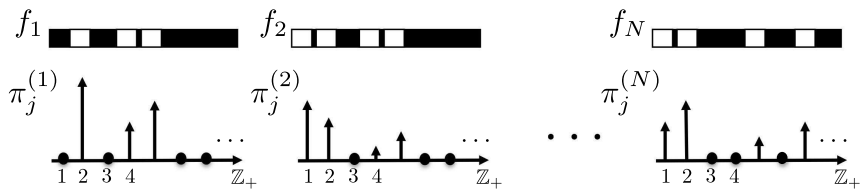}

\caption{Illustration of generating \emph{feature-constrained}
transition distributions $\pi_j^{(i)}$. Each time series' binary
feature vector $\mathbf{f}_{ i}$ limits the support of the transition
distribution to the sparse set of \mbox{selected} dynamic behaviors. The
nonzero components are Dirichlet distributed, as described by
equation~(\protect\ref{eqnDirPrior2}). The feature vectors are as in
Figure~\protect\ref{figGroundTruthMocap6}.} \label{figfeatconsttransdist}\vspace*{-3pt}
\end{figure}

The feature vectors are used to define a set of \emph
{feature-constrained transition distributions} that restrict each time
series $i$ to only switch between its set of selected behaviors, as
indicated by $\mathbf{f}_{ i}$. Let ${\pi}_{k}^{(i)}$ denote
the feature-constrained transition distribution from state $k$ for time
series $i$. Then, ${\pi}_{k}^{(i)}$ satisfies $\sum_j
{\pi}_{kj}^{(i)} = 1$, and
%
%
\begin{eqnarray}
\cases{ {\pi}_{kj}^{(i)} = 0, &\quad if $f_{ij} =
0$,
\vspace*{3pt}\cr
{\pi}_{kj}^{(i)} > 0, &\quad if $f_{ij} = 1$.}
\end{eqnarray}
See Figure~\ref{figfeatconsttransdist}. Note that here we assume that
the frequency at which the time series switch between\vadjust{\goodbreak} the selected
behaviors might be time-series-specific. That is, although two actors
may both \emph{run} and \emph{walk}, they may alternate between these
behaviors in different manners.

The observations for each data stream then follow an MSP defined by the
feature-constrained transition distributions. Although the methodology
described thus far applies equally well to HMMs and other MSPs,
henceforth we focus our attention on the AR-HMM and develop the full
model specification and inference procedures needed to treat our
motivating example of visual motion capture. Specifically, let
$\symsubsupB{y}{t}{i}$ represent the observed value of the $i$th time
series at time $t$, and let ${z}_{t}^{(i)}$ denote the latent
dynamical state. Assuming an order-$r$ AR-HMM as defined in
equation~(\ref{eqnSVAR}), we have
%
%
\begin{eqnarray}\label{eqnmultSVAR}
{z}_{t}^{(i)} \mid{z}_{t-1}^{(i)} &\sim& {\pi
}_{{z}_{t-1}^{(i)}}^{(i)},
\nonumber\\[-8pt]\\[-8pt]
\symsubsupB{y} {t} {i} \mid{z}_{t}^{(i)} & \sim&\mathcal{N}\bigl(
\mathbf{A}_{{z}_{t}^{(i)}}\symsubsupB{\tilde {y}} {t} {i}, \Sigma_{{z}_{t}^{(i)}} \bigr).\nonumber
\end{eqnarray}
Conditioned on the set of feature vectors, $\mathbf{f}_{ i}$, for
$i=1,\ldots,N$, the model reduces to a collection of $N$ switching VAR
processes, each defined on the finite state space formed by the set of
selected behaviors for that time series. The dynamic behaviors $\theta
_k = \{\mathbf{A}_{k},\Sigma_k\}$ are shared across all time series. The
feature-constrained transition distributions $\pi_j^{(i)}$ restrict
time series $i$ to select among the dynamic behaviors available in its
feature vector $\mathbf{f}_{ i}$. Each time step $t$ is assigned to
one behavior, according to assignment variable ${z}_{t}^{(i)}$.

This proposed featural model has several advantages. By discovering the
pattern of behavior sharing (i.e., discovering $f_{ik}=f_{jk}=1$ for
some pair of sequences~$i,j$), we can interpret how the time series
relate to one another. \mbox{Additionally}, behavior-sharing allows multiple
sequences to pool observations from the same behavior, improving
estimates of $\theta_k$.
%
\section{Prior specification}\label{secprior}
To maintain an unbounded set of possible behaviors, we take a Bayesian
nonparametric approach and define a model for a globally shared set of
\emph{infinitely} many possible dynamic behaviors. We first explore a
prior specification for the corresponding infinite-dimensional feature
vectors $\mathbf{f}_{ i}$. We then address the challenge of defining a
prior on infinite-dimensional transition distributions with support
constraints defined by the feature vectors.
%
\subsection{Feature vectors}\label{secfeatureprior}
Inferring the structure of behavior sharing within a Bayesian framework
requires defining a prior on the feature inclusion probabilities. Since
we want to maintain an unbounded set of possible behaviors (and thus
require infinite-dimensional feature vectors), we appeal to a Bayesian
nonparametric featural model based on the \emph{beta process-Bernoulli
process}. Informally, one can think of the formulation in our case as
follows. A beta process (BP) random measure, $B = \sum_k \omega
_k\delta_{\theta_k}$, defines an infinite set of coin-flipping
probabilities $\omega_k$---one for each behavior $\theta_k$. Each
time series $i$ is associated with a Bernoulli process realization,
$X_i = \sum_k f_{ik}\delta_{\theta_k}$, that is the outcome of an
infinite coin-flipping sequence based on the BP-determined coin
weights. The set of resulting \emph{heads} ($f_{ik}=1$) indicates the
set of selected \emph{behaviors}, and implicitly defines an
infinite-dimensional \emph{feature vector} $\mathbf{f}_{ i}$.

The properties of the BP induce sparsity in the feature space by
encouraging sharing of features among the Bernoulli process
realizations. Specifically, the total sum of coin weights is finite,
and only certain behaviors have large coin weights. Thus, certain
features are more prevalent, although feature vectors clearly need not
be identical. As such, this model allows infinitely many possible
behaviors, while encouraging a sparse, finite representation and
flexible sharing among time series. The inherent conjugacy of the BP to
the Bernoulli process allows for an analytic predictive distribution
for a feature vector based on the feature vectors observed so far. As
outlined in Section~\ref{secIBP}, this predictive distribution can be
described via the Indian buffet process~[\citet{Ghahramani06}]
under certain parameterizations of the BP. Computationally, this
representation is key.
\subsubsection*{The beta process---Bernoulli process featural model}
The BP is a special case of a general class of stochastic processes
known as~\emph{completely random measures}~[\citet{Kin1967}]. A
completely random measure $B$ is defined such that for any disjoint
sets $A_1$ and $A_2$ (of some sigma algebra $\mathcal{A}$ on a
measurable space~$\Theta$), the corresponding random variables
$B(A_1)$ and $B(A_2)$ are independent. This idea generalizes the family
of \emph{independent increments processes} on the real line. All
completely random measures can be constructed from realizations of a
nonhomogenous Poisson process [up to a deterministic component;
see~\citet{Kin1967}]. Specifically, a Poisson rate measure $\nu$
is defined on a product space $\Theta\otimes\mathbb{R}$, and a draw
from the specified Poisson process yields a collection of points $\{
\theta_j,\omega_j\}$ that can be used to define a completely random measure:
%
%
\begin{equation}
B = \sum_{k=1}^\infty\omega_k
\delta_{\theta_k}. \label{eqnCRM}
\end{equation}
This construction assumes $\nu$ has infinite mass, yielding a
countably infinite collection of points from the Poisson process.
Equation~(\ref{eqnCRM}) shows that completely random measures are
discrete. Consider a rate measure defined as the product of an
arbitrary sigma-finite \emph{base measure} $B_0$, with total mass
$B_0(\Theta)=\alpha$, and an improper beta distribution on the
interval $[0,1]$. That is, on the product space $\Theta\otimes[0,1]$
we have the following rate measure:
%
%
\begin{equation}
\nu(d\omega, d\theta) = c\omega^{-1}(1 - \omega)^{c-1}\,d\omega\,
B_0(d\theta),
\end{equation}
where $c>0$ is referred to as a \emph{concentration parameter}. The
resulting completely random measure is known as the \emph{beta
process}, with draws denoted by $B \sim\mbox{BP}(c,B_0)$. 
With this construction, the weights $\omega_k$ of the atoms in $B$ lie
in the interval $(0,1)$, thus defining our desired feature-inclusion
probabilities. 

The BP is conjugate to a class of \emph{Bernoulli
processes}~[\citet{Thibaux07}], denoted by $\mbox{BeP}(B)$,
which provide our desired feature representation. A realization
%
%
\begin{equation}
X_i\mid B \sim\mbox{BeP}(B),
\end{equation}
with $B$ an atomic measure, is a collection of unit-mass atoms on
$\Theta$ located at some subset of the atoms in $B$. In particular,
$f_{ik} \sim\operatorname{Bernoulli}(\omega_k)$ is sampled independently for
each atom $\theta_k$ in $B$, and then
%
%
\begin{equation}
X_i = \sum_k f_{ik}
\delta_{\theta_k}. \label{eqnBeP}
\end{equation}
%
One can visualize this process as walking along the atoms of a discrete
measure $B$ and, at each atom $\theta_k$, flipping a coin with
probability of heads given by $\omega_k$. Since the rate measure $\nu
$ is $\sigma$-finite, Campbell's theorem~[\citet{Kingman93}]
guarantees that for $\alpha$ finite, $B$ has finite expected measure
resulting in a finite set of ``heads'' (active features) in each $X_i$.

%
%


Computationally, Bernoulli process realizations $X_i$ are often
summarized by an infinite vector of binary indicator variables $\mathbf
{f}_{ i} = [f_{i1}, f_{i2}, \ldots]$.
Using the BP measure $B$ to tie together the feature vectors encourages
the $X_i$ to share similar features while still allowing significant
variability. 

\subsection{Feature-constrained transition distributions}
We seek a prior for transition distributions $\mbox{$\bolds{\pi}^{(i)}
=\{{\pi}_{k}^{(i)}\}$}$ defined on an infinite-dimensional
state space, but with positive support restricted to a finite subset
specified by $\mathbf{f}_{ i}$. Motivated by the fact that
Dirichlet-distributed probability mass functions can be generated via
normalized gamma random variables, for each time series $i$ we define a
doubly-infinite collection of random variables:
%
%
\begin{equation}
{\eta}_{jk}^{(i)}\mid\gamma,\kappa\sim\operatorname{Gamma}\bigl(\gamma
+\kappa\delta(j,k),1\bigr). \label{eqntransitionGamma}
\end{equation}
Here, the Kronecker delta function is defined by $\delta(j,k)=0$ when
$j \neq k$ and $\delta(k,k)=1$. The hyperparameters $\gamma,\kappa$
govern Markovian state switching\vspace*{1pt} probabilities. Using this collection
of \emph{transition weight} variables, denoted by $\bolds{\eta
}^{(i)}$, we define time-series-specific, feature-constrained
transition distributions:
%
%
\begin{eqnarray}
{\pi}_{j}^{(i)} = \frac{
[\matrix{
{\eta}_{j1}^{(i)} & {\eta}_{j2}^{(i)} & \cdots
}]
\odot\mathbf{f}_{ i}}{\sum_{k|f_{ik}=1} {\eta}_{jk}^{(i)}}, \label{eqnnormEta}
\end{eqnarray}
where $\odot$ denotes the element-wise, or Hadamard, vector product.
This construction defines ${\pi}_{j}^{(i)}$ over the full\vspace*{1pt} set
of positive integers, but assigns positive mass only at indices~$k$
where $f_{ik}=1$, constraining time series $i$ to only transition among
behaviors indicated by its feature vector $\mathbf{f}_{ i}$. See
Figure~\ref{figfeatconsttransdist}.

The preceding generative process can be equivalently represented via a
sample ${\tilde{\pi}}_{j}^{(i)}$ from a finite Dirichlet
distribution of dimension $K_i = \sum_k f_{ik}$, containing the
nonzero entries of ${\pi}_{j}^{(i)}$:
%
%
\begin{equation}
{\tilde{\pi}}_{j}^{(i)} \mid\mathbf{f}_{ i},
\gamma,\kappa \sim\operatorname{Dir}\bigl([\gamma,\ldots,\gamma,\gamma+ \kappa,\gamma,
\ldots,\gamma]\bigr). \label{eqnDirPrior}
\end{equation}
This construction reveals that $\kappa$ places extra expected mass on
the self-transition probability of each state, analogously to the
sticky HDP-HMM~[\citet{FoxAOAS11}]. We also use the representation
%
%
\begin{equation}
{\pi}_{j}^{(i)} \mid\mathbf{f}_{ i},\gamma,\kappa
\sim\operatorname{Dir}\bigl([\gamma,\ldots,\gamma,\gamma+ \kappa,\gamma,\ldots]\odot
\mathbf{f}_{ i}\bigr),\label{eqnDirPrior2}
\end{equation}
implying ${\pi}_{j}^{(i)} =
[\matrix{
{\pi}_{j1}^{(i)} & {\pi}_{j2}^{(i)} & \cdots
}]$ has only a finite number of nonzero entries~${\pi }_{jk}^{(i)}$.
This representation is an abuse of notation since the
Dirichlet distribution is not defined for infinitely many parameters.
However, the notation of equation~(\ref{eqnDirPrior2}) is useful in
reminding the reader that the indices of ${\tilde{\pi }}_{j}^{(i)}$
defined by equation~(\ref{eqnDirPrior}) are not over 1 to
$K_i$, but rather over the $K_i$ values of $k$ such that $f_{ik}=1$.
Additionally, this notation is useful for concise representations of
the posterior distribution.


We construct the model using the unnormalized transition weights
$\bolds{\eta}^{(i)}$ instead of just the proper distributions $\bolds
{\pi}^{(i)}$ so that we may consider adding or removing states when
sampling from the nonparametric posterior. Working with $\bolds{\eta
}^{(i)}$ here simplifies expressions, since we need not worry about the
normalization constraint required with $\bolds{\pi}^{(i)}$.


\subsection{VAR parameters}
%
To complete the Bayesian model specification, a conjugate matrix-normal
inverse-Wishart (MNIW) prior [cf.,~\citet{West}] is placed on the
shared collection of dynamic parameters $\theta_k = \{\mathbf
{A}_k,\Sigma
_k\}$. Specifically, this prior is comprised of an inverse Wishart
prior on $\Sigma_k$ and (conditionally) a matrix normal prior on
$\mathbf{A}_k$:
%
%
\begin{eqnarray} \label{eqnMNIW}
\Sigma_k \mid n_0,S_0 &\sim&
\mbox{IW}(n_0,S_0),
\nonumber\\[-8pt]\\[-8pt]
\mathbf{A}_k \mid\Sigma_k,M,L &\sim&\mathcal{M}
\mathcal{N} (\mathbf{A}_k;M,\Sigma_k,L),\nonumber
\end{eqnarray}
with $n_0$ the degrees of freedom, $S_0$ the scale matrix, $M$ the mean
dynamic matrix, and $L$ a matrix that together with $\Sigma_k$ defines
the covariance of $A_k$. This prior defines the base measure $B_0$ up
to the total mass parameter $\alpha$, which has to be separately
assigned (see Section~\ref{secIBPhyperparameters}). The MNIW density
function is provided in the supplemental article~[\citet{BPARHMMSupp}].

%
%
\begin{figure}

\includegraphics{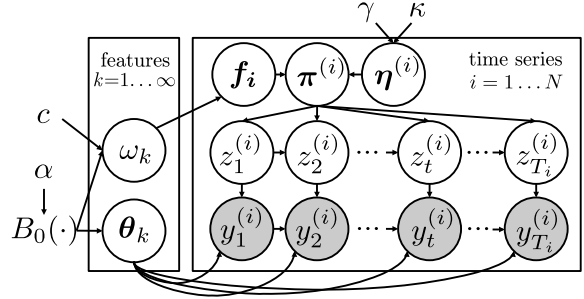}

\caption
{Graphical model representation of the BP-AR-HMM. For clarity, the
feature-inclusion probabilities, $\omega_k$, and VAR parameters,
$\theta_k$, of the beta process base measure $B\sim\mbox{BP}(c,B_0)$
are decoupled. Likewise, the Bernoulli process realizations $X_i$
associated with each time series are compactly represented in terms of
feature vectors $\mathbf{f}_{ i}$ indexed over the $\theta_k$; here,
$\mbox{$f_{ik} \mid\omega_k \sim\operatorname{Bernoulli}(\omega_k)$}$.
See equation~(\protect\ref{eqnCRM}) and equation~(\protect\ref{eqnBeP}). The $\mathbf{f}_{ i}$ are used to define
feature-constrained transition distributions $\mbox{${\pi }_{j}^{(i)}
\mid\mathbf{f}_{ i} \sim\operatorname{Dir}([\gamma,\ldots,\gamma,\gamma+\kappa,\gamma,\ldots]\odot\mathbf{f}_{ i})$}$. $\bolds
{\pi }^{(i)}$ can also be written in terms of transition weights $\bolds
{\eta}^{(i)}$, as in equation~(\protect\ref{eqnnormEta}). The state
evolves as ${z}_{t}^{(i)} \mid{z}_{t}^{(i)} \sim
{\pi}_{{z}_{t}^{(i)}}^{(i)}$ and defines conditionally\vspace*{-5pt}
VAR dynamics for $\symsubsupB{y}{t}{i}$ as in equation~(\protect\ref{eqnmultSVAR}).}\label{figBPARHMM}
\end{figure}

\section{Model overview}\label{secoverview}
Our beta-process-based featural model couples the dynamic behaviors
exhibited by different time series. We term the resulting model the
\emph{BP-autoregressive-HMM} (BP-AR-HMM). Figure~\ref{figBPARHMM}
provides a graphical model representation. Considering the \emph
{feature space} (i.e., set of autoregressive parameters) and the \emph
{temporal dynamics} (i.e., set of transition distributions) as separate
dimensions, one can think of the BP-AR-HMM as a spatio-temporal process
comprised of a (continuous) beta process in space and discrete-time
Markovian dynamics in time. The overall model specification is
summarized as follows:
\begin{enumerate}[(3)]
\item[(1)] Draw beta process realization $B \sim\mbox{BP}(c,B_0)$:
\[
B = \sum_{k=1}^{\infty} \omega_{k}
\theta_k\qquad\mbox{where } \theta_k = \{
\mathbf{A}_k, \Sigma_k\}.
\]
\item[(2)] For each sequence $i$ from $1$ to $N$:
\begin{enumerate}[(a)]
\item[(a)] Draw feature vector $\mathbf{f}_{ i} \mid B \sim\mbox{BeP}(B)$.

\item[(b)] Draw feature-constrained transition distributions
\[
{\pi}_{j}^{(i)} \mid\mathbf{f}_{ i}
\sim\operatorname{Dir}\bigl(\bigl[ \ldots, \gamma+ \delta(j,k) \kappa, \ldots\bigr]\odot
\mathbf{f}_{ i}\bigr).
\]

\item[(c)] For each time step $t$ from $1$ to $T_i$:
\begin{enumerate}[(ii)]
\item[(i)] Draw state sequence ${z}_{t}^{(i)} \mid{z}_{t-1}^{(i)} \sim{\pi
}_{{z}_{t-1}^{(i)}}^{(i)}$.

\item[(ii)] Draw observations $\symsubsupB{y}{t}{i} \mid{z}_{t}^{(i)} \sim
\mathcal{N}( \mathbf{A}_{{z}_{t}^{(i)}}\symsubsupB
{\tilde{y}}{t}{i},  \Sigma_{{z}_{t}^{(i)}})$.
\end{enumerate}
\end{enumerate}
\end{enumerate}
One can also straightforwardly consider conditionally independent
emissions in place of the VAR processes, resulting in a \emph{BP-HMM} model.




\section{MCMC posterior computations} \label{secMCMC}
In this section we develop an MCMC algorithm which aims to produce
posterior samples of the discrete indicator variables (binary feature
assignments $\mathbf{F}=\{\mathbf{f}_{ i}\}$ and state sequences
$\mathbf
{z} = \{\mathbf{z}^{(i)}\}$) underlying the BP-AR-HMM. We analytically
\emph{marginalize} the continuous emission parameters $\bolds{\theta
}=\{\mathbf{A}_k,\Sigma_k\}$ and transition weights $\bolds{\eta}=\{
\bolds{\eta}^{(i)}\}$, since both have conditionally conjugate
priors. This focus on discrete parameters represents a major departure
from the samplers developed by~\citet{BPHMM} and~\citet
{BPHMMNIPS12}, which explicitly sampled continuous parameters and
viewed $\mathbf{z}$ as auxiliary variables.

Our focus on the discrete latent structure has several benefits. First,
fixed feature assignments $\mathbf{F}$ instantiate a set of \emph{finite}
AR-HMMs, so that dynamic programming can be used to efficiently compute
marginal likelihoods. Second, we can tractably compute the joint
probability of $(\mathbf{F}, \mathbf{z}, \mathbf{y})$, which allows
meaningful comparison of configurations $(\mathbf{F},\mathbf{z})$ with
varying numbers $K_+$ of active features. Such comparison is not
possible when instantiating $\bolds{\theta}$ or $\bolds{\eta}$,
since these variables have dimension proportional to $K_+$. Finally,
our novel split-merge and data-driven birth moves both consider adding
new behaviors to the model, and we find that proposals for
fixed-dimension discrete variables are much more likely to be accepted
than proposals for high-dimensional continuous parameters. Split-merge
proposals with high acceptance rates are essential to the experimental
successes of our method, since they allow potentially large changes at
each iteration.

At each iteration, we cycle among seven distinct sampler moves:
\begin{longlist}[(3)]
\item[(1)] (Section~\ref{secsampleEtaTheta}) Sample behavior-specific
auxiliary variables: $\bolds{\theta},\bolds{\eta} \mid\mathbf{F},
\mathbf{z}$.

\item[(2)] (Section~\ref{secfeatureSampling}) Sample \emph{shared}
features, collapsing state sequences: $\mathbf{F} \mid\bolds{\theta},
\bolds{\eta}$.

\item[(3)] (Section~\ref{seczSampling}) Sample each state sequence:
$\mathbf{z} \mid\mathbf{F}, \bolds{\theta}, \bolds{\eta}$.

\item[(4)] (Section~\ref{secIBPhyperparameters}) Sample BP
hyperparameters: $\alpha,c \mid\mathbf{F}$.

\item[(5)](Section~\ref{secIBPhyperparameters}) Sample HMM transition
hyperparameters: $\gamma,\kappa\mid\mathbf{F},\bolds{\eta}$.

\item[(6)] (Section~\ref{seczDD}) Propose birth/death moves on joint
configuration: $\mathbf{F},\mathbf{z}$.

\item[(7)] (Section~\ref{secSplitMerge}) Propose split/merge move on joint
configuration: $\mathbf{F},\mathbf{z}$.
\end{longlist}
Note that some moves instantiate $\bolds{\theta},\bolds{\eta}$ as
\emph{auxiliary variables} to make computations tractable and block
sampling possible. However, we discard these variables after step 5 and
only propagate the core state space $(\mathbf{F}, \mathbf{z}, \alpha,c,
\gamma,\kappa)$ across iterations. Note also that steps 2--3 comprise
a block sampling of $\mathbf{F},\mathbf{z}$. Our MCMC steps are detailed
in the remainder of this section, except for split-merge moves which
are discussed in Section~\ref{secSplitMerge}. Further information for
all moves is also available in the supplemental article~[\citet{BPARHMMSupp}], including a summary of the overall MCMC procedure in
Algorithm~D.1.

\subsection*{Computational complexity} The most expensive step of our
sampler occurs when sampling the entries of $\mathbf{F}$ (step~2).
Sampling each binary entry requires one run of the forward--backward
algorithm to compute the likelihood $ p(\symsubsupB{y}{1\dvtx T_i}{i}\mid
\mathbf{f}_{ i}, \bolds{\eta}^{(i)},  \bolds{\theta})$; this
dynamic\vspace*{1pt} programming routine has complexity $\mathcal{O}(T_i K_i^2)$,
where $K_i$ is the number of active behavior states in sequence $i$ and
$T_i$ is the number of time steps. Computation may be significantly
reduced by caching the results of some previous sampling steps, but
this remains the most costly step. Resampling the $N$ state sequences
$\mathbf{z}$ (step~3) also requires an $\mathcal{O}(T_i K_i^2)$
forward--backward routine, but harnesses computations made in sampling
$\mathbf{F}$ and is only performed $N$ times rather than $NK$, where $K$
is the total number of instantiated features. The birth/death moves
(step~6) basically only involve the computational cost of sampling the
state sequences. Split-merge moves (step~7) are slightly more complex,
but again primarily result in repeated resampling of state sequences.
Note that although each iteration is fairly costly, the sophisticated
sampling updates developed in the following sections mean that fewer
iterations are needed to achieve reasonable posterior estimates.

Conditioned on the set of instantiated features $\mathbf{F}$ and behaviors
$\bolds{\theta}$, the model reduces to a collection of independent,
finite AR-HMMs. This structure could be harnessed to distribute
computation, and parallelization of our sampling scheme is a promising
area for future research.



%
\subsection{Background: The Indian buffet process}\label{secIBP}
Sampling the features $\mathbf{F}$ requires some prerequisite knowledge.
As shown by~\citet{Thibaux07}, marginalizing over the latent beta
process $B$ in the beta process-Bernoulli process hierarchy and taking
$c=1$ induces a predictive distribution on feature indicators known as
the Indian buffet process (IBP)~[\citet{Ghahramani06}].\footnote
{Allowing any $c>0$ induces a two-parameter IBP with a similar
construction.} The IBP is based on a culinary metaphor in which
customers arrive at an infinitely long buffet line of dishes (\emph
{features}). The first arriving customer (\emph{time series}) chooses
$\operatorname{Poisson}(\alpha)$ dishes. Each subsequent customer~$i$ selects
a previously tasted dish~$k$ with probability $m_k/i$ proportional to
the number of previous customers $m_k$ to sample it, and also samples
$\operatorname{Poisson}(\alpha/i)$ new dishes. 

For a detailed derivation of the IBP from the beta process-Bernoulli
process formulation of Section~\ref{secfeatureprior}, see
Supplement~A of~\citet{BPARHMMSupp}.
%
\subsection{Sampling shared feature assignments} \label{secfeatureSampling}
We now consider sampling each sequence's binary feature assignment
$\mathbf{f}_{ i}$. Let $\mathbf{F}^{-ik}$ denote the set of all feature
indicators excluding $f_{ik}$, and $K_+^{-i}$ be the number of
behaviors used by all other time series. Some of the $K_+^{-i}$
features may also be shared by time series $i$, but those unique to
this series are not included. For simplicity, we assume that these
behaviors are indexed by $\{1,\ldots,K_+^{-i}\}$. The IBP prior
differentiates between this set of ``shared'' features that other time
series have already selected and those ``unique'' to the current
sequence and appearing nowhere else. We may safely alter sequence $i$'s
assignments to shared features $\{1,\ldots,K_+^{-i}\}$ without changing
the number of behaviors present in $\mathbf{F}$. We give a procedure for
sampling these entries below. Sampling unique features requires adding
or deleting features, which we cover in Section~\ref{seczDD}.

Given observed data $\symsubsupB{y}{1\dvtx T_i}{i}$, transition variables
$\bolds{\eta}^{(i)}$, and emission parameters $\bolds{\theta}$, the
feature indicators $f_{ik}$ for the $i$th sequence's shared features $k
\in\{1,\ldots,K_+^{-i}\}$ have posterior distribution
%
%
\begin{equation}
p\bigl(f_{ik}\mid\mathbf{F}^{-ik}, \symsubsupB{y}
{1\dvtx T_i} {i},\bolds{\eta }^{(i)},  \bolds{\theta}\bigr)
\propto p\bigl(f_{ik}\mid\mathbf{F}^{-ik} \bigr) p\bigl(
\symsubsupB{y} {1\dvtx T_i} {i}\mid \mathbf {f}_{ i}, \bolds{
\eta}^{(i)},  \bolds{\theta}\bigr). \label{eqnFsampling}
\end{equation}
Here, the IBP prior implies that $p(f_{ik}=1\mid\mathbf{F}^{-ik}) =
m_k^{-i}/N$, where $m_k^{-i}$ denotes the number of sequences \emph
{other} than $i$ possessing $k$. This exploits the exchangeability of
the IBP~[\citet{Ghahramani06}], which follows from the BP
construction~[\citet{Thibaux07}].

When sampling binary indicators like $f_{ik}$, Metropolis--Hastings
proposals can mix faster~[\citet{Frigessi93}] and have greater
efficiency~[\citet{Liu96}] than standard Gibbs samplers. To
update $f_{ik}$ given $\mathbf{F}^{-ik} $, we thus use equation~(\ref
{eqnFsampling}) to evaluate a Metropolis--Hastings proposal which flips
$f_{ik}$ to the binary complement $\bar{f}=1-f$ of its current value $f$:
%
%
\begin{eqnarray}\label{eqnsharedFeaturesMH}
f_{ik} &\sim&\rho(\bar{f} \mid f)\delta(f_{ik},\bar{f}) +
\bigl(1-\rho (\bar{f} \mid f)\bigr)\delta(f_{ik},f),
\nonumber\\[-8pt]\\[-8pt]
\rho(\bar{f} \mid f) &=& \min \biggl\{\frac{p(f_{ik}=\bar{f}\mid
\mathbf{F}^{-ik}, \symsubsupB{y}{1\dvtx T_i}{i},\bolds{\eta}^{(i)},\theta
_{1\dvtx K_+^{-i}},c)}{p(f_{ik}=f\mid\mathbf{F}^{-ik}, \symsubsupB
{y}{1\dvtx T_i}{i},\bolds{\eta}^{(i)}, \theta_{1\dvtx K_+^{-i}},c)},1 \biggr\}.\nonumber
\end{eqnarray}
To compute likelihoods $p(\symsubsupB{y}{1\dvtx T_i}{i}\mid\mathbf{f}_{
i}, \bolds{\eta}^{(i)},  \bolds{\theta})$, we combine $\mathbf
{f}_{ i}$ and $\bolds{\eta}^{(i)}$ to construct the transition
distributions ${\pi}_{j}^{(i)}$ as in equation~(\ref
{eqnnormEta}), and marginalize over the possible latent state
sequences by applying a forward--backward message passing algorithm for
AR-HMMs [see Supplement~C.2 of~\citet{BPARHMMSupp}].
In each sampler iteration, we apply these proposals sequentially to
each entry of the feature matrix $\mathbf{F}$, visiting each entry one at
a time and retaining any accepted proposals to be used as the fixed
$\mathbf{F}^{-ik}$ for subsequent proposals. 

%
\subsection{Sampling state sequences ${z}$}\label{seczSampling}
For each sequence $i$ contained in $\mathbf{z}$, we block sample
$\mathbf{z}^{(i)}_{1\dvtx T_i}$ in one coherent move. This\vspace*{1pt} is possible
because $\mathbf{f}_{ i}$ defines a finite AR-HMM for each sequence,
enabling dynamic programming with auxiliary variables $\bolds{\pi
}^{(i)}, \bolds{\theta}$. 
We compute backward messages $m_{t+1,t}({z}_{t}^{(i)}) \propto
p(\symsubsupB{y}{t+1\dvtx T_i}{i} \mid{z}_{t}^{(i)},\symsubsupB
{\tilde{y}}{t}{i},\bolds{\pi}^{(i)},\bolds{\theta})$, and recursively
sample each ${z}_{t}^{(i)}$:
%
%
\begin{equation}
\qquad {z}_{t}^{(i)} \mid{z}_{t-1}^{(i)},
\symsubsupB {y} {1\dvtx T_i} {i},\bolds{\pi}^{(i)},  \bolds{
\theta} \sim{\pi }_{{z}_{t-1}^{(i)}}^{(i)} \bigl({z}_{t}^{(i)}
\bigr) \mathcal{N} \bigl(\symsubsupB{y} {t} {i}; \mathbf{A}_{{z}_{t}^{(i)}}\symsubsupB {
\tilde{y}} {t} {i}, \Sigma_{{z}_{t}^{(i)}} \bigr) m_{t+1,t}\bigl({z}_{t}^{(i)}
\bigr). \label{eqnzSampling}
\end{equation}
Supplement Algorithm~D.3 of~\citet
{BPARHMMSupp} explains backward-filtering, forward-sampling in detail.

%

\subsection{Sampling auxiliary parameters: \texorpdfstring{${\theta}$}{theta} and \texorpdfstring{${\eta}$}{eta}}\label{secsampleEtaTheta}
Given fixed features $\mathbf{F}$ and state sequences $\mathbf{z}$, the
posterior over auxiliary parameters factorizes neatly:
%
%
\begin{equation}
p( \bolds{\theta}, \bolds{\eta} \mid\mathbf{F}, \mathbf{z}, \mathbf{y}) = \prod
_{k=1}^{K_+} p\bigl( \theta_k
\mid\bigl\{ \mathbf{y}^{(i)}_t\dvtx   {z}_{t}^{(i)}
= k\bigr\} \bigr) \prod_{i=1}^N p\bigl(
\bolds{\eta}^{(i)} \mid\mathbf{z}^{(i)}, \mathbf{f}_{ i}
\bigr).
\end{equation}
We can thus sample each $\theta_k$ and $\bolds{\eta}^{(i)}$
independently, as outlined below.

\subsubsection*{Transition weights ${\eta}^{(i)}$}
Given state sequence $\mathbf{z}^{(i)}$ and features $\mathbf{f}_{
i}$, sequence $i$'s Markov transition weights $\bolds{\eta}^{(i)}$
have posterior distribution
%
%
\begin{equation}
p\bigl( {\eta}_{jk}^{(i)} \mid\mathbf{z}^{(i)},
f_{ij}=1, f_{ik}=1 \bigr) \propto\frac{ ( {\eta}_{jk}^{(i)})^{ {n}_{jk}^{(i)}
+ \gamma+ \kappa\delta(j,k) -1 } e^{ - {\eta}_{jk}^{(i)} }}{
 [\sum_{k'\dvtx  f_{ik'}=1} {\eta}_{jk'}^{(i)}  ]^{
{n}_{j}^{(i)} } },
\label{eqnetaPosterior}
\end{equation}
where ${n}_{jk}^{(i)}$ counts\vspace*{2pt} the transitions from state $j$ to
$k$ in ${z}_{1\dvtx {T_i}}^{(i)}$, and $n_{j}^{(i)} = \sum_k
{n}_{jk}^{(i)}$ counts all transitions out of state $j$.

Although the posterior in equation~(\ref{eqnetaPosterior}) does not
belong to any standard parametric family, simulating posterior draws is
straightforward. We use a simple \mbox{auxiliary} variable method which
inverts the usual gamma-to-Dirichlet scaling transformation used to
sample Dirichlet random variables. We explicitly draw ${\pi
}_{j}^{(i)}$, the normalized transition probabilities out of state $j$, as
%
%
\begin{eqnarray}
{\pi}_{j}^{(i)} \mid
\mathbf{z}^{(i)} &\sim&\operatorname{Dir}\bigl(\bigl[\ldots,\gamma+
{n}_{jk}^{(i)} + \kappa\delta(j,k),\ldots\bigr]\odot
\mathbf{f}_{ i}\bigr).\label{eqnpiPosterior}
\end{eqnarray}
The unnormalized transition parameters ${\eta}_{j}^{(i)}$ are
then given by the deterministic transformation ${\eta }_{j}^{(i)} =
{C}_{j}^{(i)} {\pi}_{j}^{(i)}$, where
%
%
\begin{eqnarray}
{C}_{j}^{(i)} &\sim&
\operatorname{Gamma}\bigl( K_{ +}^{(i)} \gamma + \kappa, 1
\bigr). \label{eqnetaScale}
\end{eqnarray}
Here, $K_{ +}^{(i)} = \sum_k f_{ik}$. This sampling process ensures
that transition weights $\bolds{\eta}^{(i)}$ have magnitude entirely
informed by the prior, while only the relative proportions are
influenced by $\mathbf{z}^{(i)}$. Note that this is a correction to
the posterior for ${\eta}_{j}^{(i)}$ presented in the earlier
work of \citet{BPHMM}.

%
\subsubsection*{Emission parameters $\theta_k$}
The emission
parameters $\theta_k=\{\mathbf{A}_k,\Sigma_k\}$ for each feature~$k$
have the conjugate matrix normal inverse-Wishart (MNIW) prior of
equation~(\ref{eqnMNIW}). Given $\mathbf{z}$, we form $\theta_k$'s
MNIW posterior using sufficient statistics from observations assigned
to state $k$ across all sequences $i$ and time steps $t$. Letting
$\mathbf{Y}_{ k}=\{\symsubsupB{y}{t}{i}\dvtx   {z}_{t}^{(i)} = k \}$ and
$\tilde{\mathbf{Y}}_{ k} = \{ \tilde{\mathbf{y}}_t^{(i)}:
{z}_{t}^{(i)} = k\}$, define
%
%
\begin{eqnarray}\label{eqnSk}
\qquad {S}_{\tilde{y}\tilde{y}}^{(k)} &=& \sum_{(t,i)\mid{z}_{t}^{(i)} = k}
\symsubsupB{\tilde{y}} {t} {i}\symsubsupT{\tilde {y}} {t} {i} + \mathbf{L},\qquad
{S}_{y\tilde{y}}^{(k)} = \sum
_{(t,i)\mid{z}_{t}^{(i)} = k} \symsubsupB {y} {t} {i}\symsubsupT{\tilde{y}} {t} {i} +
\mathbf{M}\mathbf{L},
\nonumber\\[-8pt]\\[-8pt]
{S}_{yy}^{(k)} &=& \sum_{(t,i)\mid{z}_{t}^{(i)} = k}
\symsubsupB{y} {t} {i}\symsubsupT{y} {t} {i} + \mathbf{M}\mathbf {L}
\mathbf{M}^T,\qquad {S}_{y|\tilde{y}}^{(k)} =
{S}_{yy}^{(k)} - {S}_{y\tilde{y}}^{(k)}S_{\tilde{y}\tilde
{y}}^{-(k)}S_{\tilde{y}\tilde{y}}^{(k)^T}.\nonumber
\end{eqnarray}
Using standard MNIW conjugacy results, the posterior is then
%
%
\begin{eqnarray}
\mathbf{A}_k \mid\Sigma_k,\mathbf{Y}_{ k},
\tilde{\mathbf{Y}}_{ k} &\sim&\mathcal{M}\mathcal{N} \bigl(
\mathbf{A}_k;{S}_{y\tilde
{y}}^{(k)}S_{\tilde{y}\tilde {y}}^{-(k)},
\Sigma_k,{S}_{\tilde
{y}\tilde{y}}^{(k)} \bigr),
\nonumber\\[-8pt]\\[-8pt]
\Sigma_k \mid\mathbf{Y}_{ k},\tilde{\mathbf{Y}}_{ k}
&\sim&\mbox{IW} \bigl(|\mathbf{Y}_{ k}| + n_0,
{S}_{y|\tilde{y}}^{(k)} + S_0 \bigr).\nonumber
\end{eqnarray}
Through sharing across multiple time series, we improve inferences
about $\{ \mathbf{A}_k, \Sigma_k \}$ compared to endowing each sequence
with separate behaviors.

\subsection{Sampling the BP and transition hyperparameters} \label{secIBPhyperparameters}

We additionally place priors on the transition hyperparameters $\gamma
$ and $\kappa$, as well as the BP parameters $\alpha$ and $c$, and
infer these via MCMC. Detailed descriptions of these sampling steps are
provided in Supplement~G.2 of~\citet{BPARHMMSupp}.
%
%

\subsection{Data-driven birth--death proposals of unique features}\label{seczDD}
We now consider exploration of the unique features associated with each
sequence. One might consider a birth--death version of a reversible jump
proposal~[\citet{Green95}] that either adds one new feature
(``birth'') or eliminates an existing unique feature. This scheme was
considered by~\citet{BPHMM}, where each proposed new feature~$k^*$ (HMM state) was associated with an emission parameter $\theta
_{k^*}$ and associated transition parameters $\{\eta^{(i)}_{jk^*},\eta
^{(i)}_{k^*j}\}$ drawn from their priors. However, such a sampling
procedure can lead to extremely low acceptance rates in
high-dimensional cases since it is unlikely that a random draw of
$\theta_{k^*}$ will better explain the data than existing,
data-informed parameters. Recall that for a BP-AR-HMM with VAR(1)
likelihoods, each $\theta_k = \{\mathbf{A}_k,\Sigma_k\}$ has
$d^2+d(d+1)/2$ scalar parameters. This issue was addressed by the
data-driven proposals of~\citet{BPHMMNIPS12}, which used randomly
selected windows of data to inform the proposal distribution for
$\theta_{k^*}$. \citet{tu02} employed a related family of
data-driven MCMC proposals for a very different image segmentation model.

The birth--death frameworks of~\citet{BPHMM} and \citet
{BPHMMNIPS12} both perform such moves by marginalizing out the state
sequence $\mathbf{z}^{(i)}$ and modifying the continuous HMM
parameters $\bolds{\theta},\bolds{\eta}$. Our proposed sampler
avoids the challenge of constructing effective proposals for $\bolds
{\theta},\bolds{\eta}$ by collapsing away these high-dimensional
parameters and only proposing modifications to the discrete assignment
variables $\mathbf{F},\mathbf{z}$, which are of fixed dimension
regardless of the dimensionality of the observations $\symsubsupB
{y}{t}{i}$. 
Our experiments in Section~\ref{secMoCap} show the improved mixing of
this discrete assignment approach over previously proposed alternative samplers.

At a high level, our birth--death moves propose changing one binary
entry in~$\mathbf{f}_{ i}$, combined with a corresponding change to
the state sequence $\mathbf{z}^{(i)}$. In particular, given a sequence
$i$ with $K_i$ active features (of which $n_i=K_i-K_+^{-i}$ are
unique), we first select the type of move (birth or death). If $n_i$ is
empty, we always propose a birth. Otherwise, we propose a birth with
probability $\frac{1}{2}$ and a death of unique feature $k$ with
probability $\frac{1}{2 n_i}$. We denote this proposal distribution as
$q_f( \mathbf{f}_i^* | \mathbf{f}_i)$. Once the feature proposal is
selected, we then propose a new state sequence configuration $\mathbf
{z}^{*(i)}$.

Efficiently drawing a proposed state sequence $\mathbf{z}^{*(i)}$
requires the backward-filtering, forward-sampling scheme of
equation~(\ref{eqnzSampling}). To perform this dynamic programming we
\emph{deterministically} instantiate the HMM transition weights and
emission parameters as auxiliary variables: $\hat{\bolds{\eta}},\hat
{\bolds{\theta}}$. These quantities are deterministic functions of
the conditioning set used solely to define the proposal and are not the
same as the actual sampled variables used, for example, in steps 1--5 of
the algorithm overview. The variables are discarded before subsequent
sampling stages. Instantiating these variables allows efficient \emph
{collapsed} proposals of discrete indicator configuration $(\mathbf{F}^*,
\mathbf{z}^*)$. To define good auxiliary variables, we harness the
data-driven ideas of~\citet{BPHMMNIPS12}. An outline is provided
below with a detailed summary and formal algorithmic presentation in
Supplement~E of~\citet{BPARHMMSupp}.

\subsubsection*{Birth proposal for ${z}^{*(i)}$} During a birth, we
create a new state sequence $\mathbf{z}^{*(i)}$ that can use any
features in $\mathbf{f}^*_i$, including the new ``birth'' feature $k^*$.
To construct this proposal, we utilize deterministic auxiliary
variables $\hat{\bolds{\theta}}, \hat{\bolds{\eta}}{}^{(i)}$. For
existing features $k$ such that $f_{ik}=1$, we set $\hat{\eta
}^{(i)}_{kj}$ to the prior mean of ${\eta}_{kj}^{(i)}$ and
$\hat{\theta}_k$ to the posterior mean of $\theta_k$ given all data
in any sequence assigned to $k$ in the current sample $\mathbf
{z}^{(i)}$. For new features $k^*$, we can similarly set $\hat{\eta
}^{(i)}_{ k^*j}$ to the prior mean. For $\hat{\theta}_{k^*}$,
however, we use a \emph{data-driven} construction since using the
vague MNIW prior mean would make this feature unlikely to explain any
data at hand.

The resulting \emph{data-driven} proposal for $\mathbf{z}^{(i)}$ is
as follows. First, we choose a random subwindow $W$ of the current
sequence $i$. $W$ contains a contiguous region of time steps within $\{
1,2,\ldots, T_i\}$. Second, conditioning on the chosen window, we set
$\hat{\theta}_{k^*}$ to the posterior mean of $\theta_k$ given the
data in the window $\{ \symsubsupB{y}{t}{i}\dvtx   t \in W\}$. Finally,
given auxiliary variables $\hat{\bolds{\theta}},\hat{\bolds{\eta
}}^{(i)}$ for \emph{all} features, not just the newborn $k^*$, we
sample the proposal $\mathbf{z}^{*(i)}$ using the efficient dynamic
programming algorithm for block-sampling state sequences. This sampling
allows \emph{any} time step in the current sequence to be assigned to
the new feature, not just those in $W$, and similarly does not force
time steps in $W$ to use the new feature $k^*$. These properties
maintain reversibility. 
We denote this proposal distribution by $q_{z\mbox{-}\mathrm{birth}}( \mathbf
{z}^* \mid\mathbf{F}^*, \mathbf{z}, \mathbf{y})$.

\subsubsection*{Death proposal for ${z}^{*(i)}$} During a death
move, we propose a new state sequence $\mathbf{z}^{*(i)}$ that only
uses the reduced set of behaviors in $\mathbf{f}^*_i$. This requires
deterministic auxiliary variables $\hat{\bolds{\theta}},\hat{\bolds
{\eta}}^{(i)}$ constructed as in the birth proposal, but here only for
the features in $\mathbf{f}^*_i$ which are all already existing. Again,
using these quantities we construct the proposed $\mathbf{z}^{*(i)}$
via block sampling and denote the proposal distribution by
$q_{z\mbox{-}\mathrm{death}}( \mathbf{z}^* \mid\mathbf{F}^*, \mathbf{z},
\mathbf{y})$.

\subsubsection*{Acceptance ratio} After constructing the proposal, we
decide to accept or reject via the Metropolis--Hastings ratio with
probability $\min(1,\rho)$, where for a birth move
%
%
\begin{equation}
\rho_{\mathrm{birth\mbox{-}in\mbox{-}seq\mbox{-}} i} = \frac{ p( \mathbf{y}, \mathbf{z}^*, \mathbf{f}_i^*) } { p( \mathbf{y},
\mathbf{z}, \mathbf{f}_i) } \frac{ q_{z\mbox{-}\mathrm{death}}( \mathbf{z} \mid\mathbf{F}, \mathbf{z}^*,
\mathbf{y}) }{
 q_{z\mbox{-}\mathrm{birth}}( \mathbf{z}^* \mid\mathbf{F}^*, \mathbf{z},
\mathbf{y}) } \frac{ q_{\mathrm{f}}( \mathbf{f}_i \mid\mathbf{f}_i^*)}{
 q_{\mathrm{f}}( \mathbf{f}^*_i \mid\mathbf{f}_i) }.
\label{eqzDDacceptRatio}
\end{equation}
Note that evaluating the joint probability $p( \mathbf{y}, \mathbf{z},
\mathbf{f}_i)$ of a new configuration in the discrete assignment space
requires considering the likelihood of \emph{all} sequences, rather
than just the current one, because under the BP-AR-HMM collapsing over~$\bolds{\theta}$ induces dependencies between all data time steps
assigned to the same feature. This is why we use notation $\mathbf
{z}$, even though our proposals only modify variables associated with
sequence $i$. The sufficient statistics required for this evaluation
are needed for many other parts of our sampler, so in practice the
additional computational cost is negligible compared to previous
birth--death approaches for the BP-AR-HMM.

Finally, we note that we need not account for the random choice of the
window $W$ in the acceptance ratio of this birth--death move. Each
possible $W$ is chosen independently of the current sampler
configuration, and each choice defines a valid transition kernel over a
reversible pair of birth--death moves satisfying detailed balance.

%
%
%
%

%

%
\section{Split-merge proposals} \label{secSplitMerge}
The MCMC algorithm presented in Section~\ref{secMCMC} defines a
correct and tractable inference scheme for the BP-AR-HMM, but the local
one-at-a-time sampling of feature assignments can lead to slow mixing
rates. In this section we propose split-merge moves that allow
efficient exploration of large changes in assignments, via simultaneous
changes to multiple sequences, and can be interleaved with the sampling
updates of Section~\ref{secMCMC}. Additionally, in Section~\ref
{secAnneal} we describe how both split-merge and birth--death moves can
be further improved via a modified annealing procedure that allows fast
mixing during sampler burn-in.

%
\subsection{Review: Split-merge for Dirichlet processes}
Split-merge MCMC\break methods for nonparametric models were first employed
by \citet{JainNeal04} in the context of Dirichlet process (DP)
mixture models with conjugate likelihoods. Conjugacy allows samplers to
operate directly on discrete partitions of observations into clusters,
marginalizing emission parameters. Jain and Neal use \emph{restricted
Gibbs} (RG) sampling to create reversible proposals that split a single
cluster $k_m$ into two ($k_a,k_b$) or merge two clusters into one.

To build an initial split, the RG sampler first assigns items
originally in cluster $k_m$ at random to either $k_a$ or $k_b$.
Starting from this partition, the sampler performs one-at-a-time Gibbs
updates, forgetting an item's current cluster and reassigning to either
$k_a$ or $k_b$ conditioned on the remaining partitioned data. A
proposed new configuration is obtained after several sweeps. 
For nonconjugate models, more sophisticated proposals are needed to
also instantiate emission parameters~[\citet{JainNeal07}].

Even in small data sets, performing many sweeps for each RG proposal is
often necessary for good performance~[\citet{JainNeal04}]. For
large data sets, however, requiring many sweeps for a single proposal
is computationally expensive. An alternative method, \emph{sequential
allocation}~[\citet{Dahl05}], replaces the random initialization
of RG. Here, two randomly chosen items ``anchor'' the initial
assignments of the two new clusters $k_a,k_b$. Remaining items are then
\emph{sequentially} assigned to either $k_a$ or $k_b$ one at a time,
using RG moves conditioning only on previously assigned data. This
creates a proposed partition 
after only one sampling sweep. Recent work has shown some success with
sequentially allocated split-merge moves for a hierarchical DP topic
model~[\citet{HDPsplitmerge}].

Beyond the DP mixture model setting, split-merge MCMC moves are not
well studied. Both \citet{Meeds07} and \citet{Morup2011}
mention adapting an RG procedure for relational models with latent
features based on the beta process. However, neither work provides
details on constructing proposals, and both lack experimental
validation that split-merge moves improve inference. 
%
\subsection{Split-merge MCMC for the BP-AR-HMM}
%
%
\begin{figure}

\includegraphics{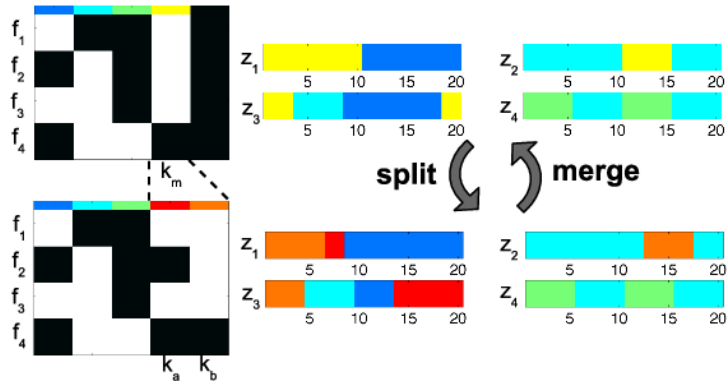}

\caption{Illustration of split-merge moves for the BP-AR-HMM, which
alter binary feature matrix $\mathbf{F}$ (white indicates present feature)
and state sequences $\mathbf{z}$. We show $\mathbf{F},\mathbf{z}$ before
(\emph{top}) and after (\emph{bottom}) feature $k_m$ (\textup{yellow}) is
split into $k_a, k_b$ (\textup{red}, \textup{orange}). An item possessing feature $k_m$
can have either $k_a,k_b$, or both after the split, and its new
$\mathbf{z}$ sequence is entirely resampled using any features
available in $\mathbf{f}_{ i}$. An item without $k_m$ cannot possess
$k_a,k_b$, and its $\mathbf{z}$ does not change. Note that a split
move can always be reversed by a merge.}\label{figSplitMergeIllustrated}
\end{figure}

In standard mixture models, such as considered by~\citet
{JainNeal04}, a given data item $i$ is associated with a single cluster
$k_i$, so selecting two anchors $i$ and $j$ is equivalent to selecting
two cluster indices $k_i,k_j$. However, in feature-based models such as
the \mbox{BP-AR-HMM}, each data item $i$ possesses a \emph{collection} of
features indicated by~$\mathbf{f}_{ i}$. Therefore, our split-merge
requires a mechanism not only for selecting anchors, but also for
choosing candidate features to split or merge from $\mathbf{f}_{
i},\mathbf{f}_{ j}$. After proposing modified feature vectors, the
associated state sequences must also be updated. Following the
motivations for our data-driven birth--death proposals, our split-merge
proposals create new feature matrices $\mathbf{F}^*$ and state sequences
$\mathbf{z}^*$, collapsing away HMM parameters $\bolds{\theta
},\bolds{\eta}$. Figure~\ref{figSplitMergeIllustrated} illustrates
$\mathbf{F}$ and $\mathbf{z}$ before and after a split proposal.
Motivated by the efficiencies of sequential allocation~[\citet
{Dahl05}], we adopt a sequential approach. Although a RG approach that
samples all variables ($\mathbf{F},\mathbf{z},\bolds{\theta},\bolds
{\eta}$) is also possible and relatively straightforward, our
experiments [Supplement~I of~\citet
{BPARHMMSupp}] show that our sequential collapsed proposals are vastly
preferred. Intuitively, constructing high acceptance rate proposals for
$\bolds{\theta}, \bolds{\eta}$ can be very difficult since each
behavior-specific parameter is high dimensional. 

\subsubsection*{Selecting anchors} Following \citet{Dahl05}, we first
select distinct anchor data items $i$ and $j$ uniformly at random from
all time series. The fixed choice of $i,j$ defines a split-merge
transition kernel satisfying detailed balance [\citet{Tierney}].
Next, we select from each anchor one feature it possesses, denoted
$k_i, k_j$, respectively. This choice determines the proposed move: we
merge $k_i,k_j$ if they are distinct, and split $k_i=k_j$ into two new
features otherwise.

Selecting $k_i,k_j$ uniformly at random is problematic. First, in data
sets with many features choosing $k_i = k_j$ is unlikely, making split
moves rare. We need to bias the selection process to consider splits
more often. Second, in a reasonably fit model most feature pairs will
not make a sensible merge. Selecting a pair that explains similar data
is crucial for efficiency. We thus develop a proposal distribution
which first draws $k_i$ uniformly from the positive entries in $\mathbf
{f}_{ i}$, and then selects $k_j$ given fixed $k_i$ as follows:
%
%
\begin{eqnarray}\label{eqnselectFeatsForSplitMerge}
q_{k}( k_i,k_j \mid\mathbf{f}_{ i},
\mathbf{f}_{ j})&=&\operatorname{Unif}\bigl( k_i \mid\{ k\dvtx  f_{ik} = 1 \} \bigr)q(k_j \mid
k_i, \mathbf{f}_{ j}),
\\
q( k_j =k \mid k_i,\mathbf{f}_j) &\propto&
\cases{ 2 R_j f_{jk}, &\quad if $k=k_i$,
\vspace*{5pt}\cr
f_{jk} \displaystyle\frac{ m( \mathbf{Y}_{k_i}, \mathbf{Y}_{k}) }{ m( \mathbf
{Y}_{k_i}) m(
\mathbf{Y}_k) }, &\quad otherwise,}
\end{eqnarray}
where $\mathbf{Y}_{k}$ denotes all observed data in any segment assigned
to $k$ (determined by~$\mathbf{z}$) and $m(\cdot)$ denotes the \emph
{marginal likelihood} of pooled data observations under the emission
distribution. A high value for the ratio $\frac{ m( \mathbf{Y}_{k_i},
\mathbf{Y}_{k}) }{ m( \mathbf{Y}_{k_i}) m( \mathbf{Y}_k) }$
indicates that the
model prefers to explain all data assigned to $k_i,k_j$ \emph
{together} rather than use a separate feature for each. This choice
biases selection toward promising merge candidates, leading to higher
acceptance rates. We set $R_j = \sum_{k_j \ne k_i} f_{jk_j} \frac{ m(
\mathbf{Y}_{k_i}, \mathbf{Y}_{k_j}) }{ m(\mathbf{Y}_{k_i})
m(\mathbf{Y}_{k_j})}$
to ensure the probability of a split (when possible) is $2/3$.

For the VAR likelihood of interest, the marginal likelihood $m( \mathbf
{Y}_k)$ of all data assigned to feature $k$, integrating over
parameters $\theta_k = \{\mathbf{A}_k,\Sigma_k\}$, is
%
%
\begin{eqnarray}
m( \mathbf{Y}_k) &=& p( \mathbf{Y}_k \mid M, L,
S_0, n_0)\nonumber
\\
&=& \int\!\!\int p( \mathbf{Y}_k \mid\mathbf{A}_k,
\Sigma_k) p( \mathbf{A}_k \mid M,\Sigma_k,
L) p( \Sigma_k \mid n_0, S_0) \,d\Sigma
_k \,d\mathbf{A}_k
\\
&=&\frac{ 1 }{ (2\pi)^{(n_k d)/2} } \cdot \frac{
\Gamma_d( (n_k+n_0)/2)
}{\Gamma_d( {n_0}/{2})
} \cdot \frac{ | S_0 |^{ {n_0}/{2} } }{
 | S^{(k)}_{y|\bar{y} } |^{ ({n_k+n_0})/{2} } }
\cdot \frac{ | L |^{{1}/{2} } }{
 | S^{(k)}_{\bar{y}\bar{y} } |^{{1}/{2} } },\nonumber
\end{eqnarray}
where $\Gamma_d(\cdot)$ is the $d$-dimensional multivariate gamma
function, $|\cdot|$ denotes the determinant, $n_k$ counts the number
of observations in set $\mathbf{Y}_k$, and sufficient statistics~$S^{(k)}_{\cdot,\cdot}$ are defined in equation~(\ref{eqnSk}).
Further details on this feature selection process are given in
Supplement~F.1, especially Algorithm~F.2, of~\citet{BPARHMMSupp}.

Once $k_i,k_j$ are fixed, we construct the candidate state $\mathbf
{F}^*,\mathbf{z}^*$ for the proposed move. This construction depends
on whether a split or merge occurs, as detailed below. Recall from
Figure~\ref{figSplitMergeIllustrated} that we only alter $\mathbf
{f}_{ \ell},\mathbf{z}^{(\ell)}$ for data sequences $\ell$ which
possess either $k_i$ or $k_j$. We call this set of items the \emph
{active set} $\mathcal{S}$. Items not in the active set are unaltered
by our proposals.

\begin{algorithm}
\caption{Construction of candidate split configuration ($\mathbf
{F},\mathbf{z}$), replacing feature $k_m$ with new features $k_a,k_b$
via sequential allocation}\label{algSplitProposal}
\fontsize{10pt}{20pt}{\selectfont{\begin{algorithmic}[1]
\STATE$f_{i,[k_a, k_b] } \gets[1 \hspace{.2cm} 0]$ \hspace{1cm} $z^{(i)}_{t\dvtx   z^{(i)}_t=k_m } \gets k_a$ \COMMENT{use anchor $i$ to create feature $k_a$}\vspace*{3pt}
\STATE$f_{j,[k_a, k_b] } \gets[0 \hspace{.2cm} 1]$ \hspace{1cm}
$z^{(j)}_{t\dvtx   z^{(j)}_t=k_m } \gets k_b$ \COMMENT{use anchor $j$ to
create feature $k_b$}
\STATE$\hat{\bolds{\theta}} \gets\mathbb{E} [ \bolds{\theta
} \mid\mathbf{y}, \mathbf{z}  ]$
[Algorithm~E.4]
\COMMENT{set emissions to posterior mean}
\STATE$\hat{\bolds{\eta}}{}^{(\ell)} \gets\mathbb{E} [ \bolds
{\eta}^{(\ell)}  ]$, $\ell\in\mathcal{S}$ [Algorithm~E.4]
\COMMENT{set transitions to prior mean}
\STATE$\mathcal{S}_{\mathrm{prev}} = \{i,j\}$ \COMMENT{initialize
set of previously visited items}
\STATE\textbf{for} nonanchor items $\ell$ in random permutation of
active set $\mathcal{S}$:
\STATE\hspace{1cm}$f_{\ell,[k_a k_b]} \sim
\cases{
[0 \hspace{.1cm} 1] \cr[1 \hspace{.1cm} 0] \propto p( f_{\ell,[k_a
k_b]} \mid F_{\mathcal{S}_{\mathrm{prev}},[k_a k_b]})p( \mathbf
{y}^{(\ell)} \mid\mathbf{f}_{ \ell}, \hat{\bolds{\theta}}, \hat
{\bolds{\eta}}^{(\ell)}) \cr[1 \hspace{.1cm} 1]}$
[Algorithm~F.4] \label{linesplitProposalFeatures}
\STATE\hspace{1cm}$\mathbf{z}^{(\ell)} \sim p(\mathbf{z}^{(\ell)}
\mid\mathbf{y}^{(\ell)}, \mathbf{f}_{ \ell}, \hat{\bolds{\theta}},
\hat{\bolds{\eta}}{}^{(\ell)})$
[Algorithm~D.3]
\STATE\hspace{1cm}\textbf{add} $\ell$ to $\mathcal{S}_{\mathrm
{prev}}$ \COMMENT{add latest sequence to set of visited items}
\STATE\hspace{1cm}\textbf{for} $k = k_a,k_b$: $\hat{\theta}_{k}
\gets\mathbb{E} [ \theta_{k} \mid\{y^{(n)}_{t}\dvtx z^{(n)}_{t} =
k, n \in\mathcal{S}_{\mathrm{prev}} \}  ]$\vspace*{5pt}
\STATE$f_{i,[k_a k_b] } \sim\cases{
[1 \hspace{.1cm} 0]
\cr[1 \hspace{.1cm} 1]}$
\hspace{.3cm}
$f_{j, [k_a k_b] } \sim\cases{
[0 \hspace{.1cm} 1]
\cr[1 \hspace{.1cm} 1]}$ \COMMENT{finish by sampling $\mathbf
{f},\mathbf{z}$ for anchors }\vspace*{3pt}
\STATE$\mathbf{z}^{(i)} \sim p( \mathbf{z}^{(i)} \mid\mathbf{y}^{(i)},
\mathbf{f}_{ i}, \hat{\bolds{\theta}}, \hat{ \bolds{\eta
}}^{(i)}$)
\hspace{.5cm} $\mathbf{z}^{(j)} \sim p( \mathbf{z}^{(j)} \mid
\mathbf{y}^{(j)}, \mathbf{f}_{ j}, \hat{\bolds{\theta}}, \hat
{\bolds
{\eta}}^{(j)}$)
\end{algorithmic}
\emph{Note: Algorithm references found in the supplemental
article~\textup{[}\citeauthor{BPARHMMSupp} \textup{(\citeyear{BPARHMMSupp})}\textup{]}}}}%
\end{algorithm}

\subsubsection*{Split} Our split proposal is defined in Algorithm~\ref
{algSplitProposal}. Iterating through a random permutation of items
$\ell$ in the active set $\mathcal{S}$, we sample $\{f^*_{\ell
k_a},f^*_{\ell k_b}\}$ from\vspace*{2pt} its conditional posterior given previously
visited items in $\mathcal{S}$, requiring that $\ell$ must possess at
least one of the new features $k_a,k_b$. We then block sample its state
sequence $\mathbf{z}^{*(\ell)}$ given $\mathbf{f}^{*}_\ell$. After
sampling all non-anchor sequences in $\mathcal{S}$, we finally sample
$\{\mathbf{f}_i^*,\mathbf{z}^{*(i)}\}$ and $\{\mathbf{f}_j^*,\mathbf
{z}^{*(j)}\}$ for anchor items $i,j$, enforcing $f^*_{ik_a}=1$ and
$f^*_{jk_b}=1$ so the move remains reversible under a merge. This does
not force $\mathbf{z}^*_i$ to use $k_a$ nor $\mathbf{z}^*_j$ to use $k_b$.

The dynamic programming recursions underlying these proposals use
nonrandom auxiliary variables in a similar manner to the data-driven
birth--death proposals. In particular, the HMM transition weights $\hat
{\bolds{\eta}}{}^{(\ell)}$ are set to the prior mean of $\bolds{\eta
}^{(\ell)}$. The HMM emission parameters $\hatBFsub{\theta}{k}$ are
set to the posterior mean of $\theta_k$ given the current data
assigned to behavior $k$ in $\mathbf{z}$ across all sequences. For new
states $k^* \in\{k_a, k_b\}$, we initialize $\hatBFsub{\theta}{k^*}$
from the anchor sequences and then update to account for new data
assigned to $k^*$ after each item $\ell$. 
As before, $\hat{\bolds{\eta}}$, $\hat{\bolds{\theta}}$ are deterministic
functions of the conditioning set used to define the \emph{collapsed}
proposals for $\mathbf{F}^*,\mathbf{z}^*$; they are discarded prior to
subsequent sampling stages.

\subsubsection*{Merge} To merge $k_a,k_b$ into a new feature $k_m$,
constructing $\mathbf{F}^*$ is deterministic: we set $f^*_{\ell k_m}=1$
for $\ell\in\mathcal{S}$, and 0 otherwise. We thus need only to
sample $\mathbf{z}^*_\ell$ for items in $\mathcal{S}$. We use a
block sampler that conditions on $\mathbf{f}_{ \ell}^*,\hat{\bolds
{\theta}}, \hat{\bolds{\eta}}{}^{(\ell)}$, where again $\hat{\bolds
{\theta}}, \hat{\bolds{\eta}}{}^{(\ell)}$ are auxiliary variables.%

\subsubsection*{Accept--reject} After drawing a candidate configuration
$(\mathbf{F}^*,\mathbf{z}^*)$, the final step is to compute a
Metropolis--Hastings acceptance ratio $\rho$. Equation~(\ref
{eqSMGenericAcceptRatio}) gives the ratio for a \emph{split} move
which creates features $k_a, k_b$ from $k_m$:
%
%
\begin{eqnarray}\label{eqSMGenericAcceptRatio}
\rho_{\mathrm{split}} &=& \frac{ p( \mathbf{y},  \mathbf{F}^*,
\mathbf
{z}^*) }{
 p( \mathbf{y}, \mathbf{F}, \mathbf{z}) } \frac{ q_{\mathrm{merge}}( \mathbf{F}, \mathbf{z} \mid\mathbf
{y},\mathbf{F}^*, \mathbf{z}^*, k_a, k_b) }{ q_{\mathrm{split}}(
\mathbf{F}^*,
\mathbf{z}^* \mid\mathbf{y},\mathbf{F}, \mathbf{z}, k_m) }
\frac{ q_{k}( k_a, k_b \mid\mathbf{y},\mathbf{F}^*, \mathbf{z}^*,
i, j)
}{ q_{k}( k_m, k_m \mid\mathbf{y},\mathbf{F}, \mathbf{z}, i, j) }. \hspace*{-25pt}
\end{eqnarray}
%
%
Recall that our sampler only updates discrete variables $\mathbf
{F},\mathbf{z}$ and marginalizes out continuous HMM parameters $\bolds
{\eta}, \bolds{\theta}$.
Our split-merge moves are therefore only tractable with conjugate
emission models such as the VAR likelihood and MNIW prior. Proposals
which instantiate emission parameters $\bolds{\theta}$, as in
\citet{JainNeal07}, would be required in the nonconjugate case.

For complete split-merge algorithmic details, consult Supplement~F of~\citet{BPARHMMSupp}. In particular, we
emphasize that the nonuniform choice of features to split or merge
requires some careful accounting, as does the correct computation of
the reverse move probabilities. These issues are discussed in the
supplemental article~[\citet{BPARHMMSupp}].

\subsection{Annealing MCMC proposals}\label{secAnneal}
We have presented two novel MCMC moves for adding or deleting features
in the BP-AR-HMM: split-merge and birth--death moves. Both propose a new
discrete variable configuration $\Psi^*=(\mathbf{F}^*, \mathbf{z}^*)$
with either one more or one fewer feature. This proposal is accepted or
rejected with probability $\min(1,\rho)$, where $\rho$ has the
generic form
%
%
\begin{equation}
\rho= \frac{ p( \mathbf{y}, \Psi^*) }{
 p( \mathbf{y}, \Psi) } \frac{ q( \Psi\mid\Psi^*, \mathbf{y}) }{
 q( \Psi^* \mid\Psi, \mathbf{y}) }.
\end{equation}
This Metropolis--Hastings ratio $\rho$ accounts for improvement in
joint probability [via the ratio of $p(\cdot)$ terms] and the
requirement of reversibility [via the ratio of $q(\cdot)$ terms]. We
call this latter ratio the \emph{Hastings factor}. Reversibility
ensures that detailed balance is satisfied, which is a sufficient
condition for convergence to the true posterior distribution.

The reversibility constraint can limit the effectiveness of our
proposal framework. Even when a proposed configuration $\Psi^*$
results in better joint probability, its Hastings factor can be small
enough to cause rejection. For example, consider any merge proposal.
Reversing this merge requires returning to the original configuration
of the feature matrix $\mathbf{F}$ via a split proposal. Ignoring anchor
sequence constraints for simplicity, split moves can produce roughly
$3^{|\mathcal{S}|}$ possible feature matrices, since each sequence in
the active set $\mathcal{S}$ could have its new features $k_a, k_b$
set to $[0~1], [1~0]$, or $[1~1]$. Returning to the exact original
feature matrix out of the many possibilities can be very unlikely. Even
though our proposals use data wisely, the vast space of possible split
configurations means the Hastings factor will always be biased toward
rejection of a merge move.

As a remedy, we recommend \emph{annealing} the Hastings factor in the
acceptance ratio of both split-merge and data-driven birth--death moves.
That is, we use a modified acceptance ratio
%
%
\begin{equation}
\rho= \frac{ p( \mathbf{y}, \Psi^*) }{
 p( \mathbf{y}, \Psi) } \biggl[ \frac{ q( \Psi\mid\Psi^*, \mathbf{y}) }{
 q( \Psi^* \mid\Psi, \mathbf{y})  } \biggr]^{ {1}/{T_{s}} },
\end{equation}
where $T_{s}$ indicates the ``temperature'' at iteration $s$. We start
with a temperature that is very large, so that $\frac{1}{T_{s}}\approx
0$ and the Hastings\vspace*{2pt} factor is ignored. The resulting greedy stochastic
search allows rapid improvement from the initial configuration. Over
many iterations, we gradually decrease the temperature toward 1. After
a specified number of iterations we fix $\frac{1}{T_{s}}=1$, so that
the Hastings factor is fully represented and the sampler is reversible.

In practice, we use an annealing schedule that linearly interpolates
$\frac{1}{T_s}$ between 0 and 1 over the first several thousand
iterations. Our experiments in Section~\ref{secMoCap} demonstrate
improvement in mixing rates based on this annealing.


\section{Related work}\label{secrelated}
Defining the number of dynamic regimes presents a challenging problem
in deploying Markov switching processes such as the \mbox{AR-HMM}. Previously,
Bayesian nonparametric approaches building on the hierarchical
Dirichlet process (HDP)~[\citet{Teh06}] have been proposed to
allow uncertainty in the number of regimes by defining Markov switching
processes on infinite state spaces~[\citet{Beal02,Teh06}, \citeauthor{FoxAOAS11} (\citeyear{FoxIEEE11,FoxAOAS11})]. See~\citet{FoxIEEESPM} for a
recent review. However, these formulations focus on a single time
series, whereas in this paper our motivation is analyzing \emph
{multiple} time series. A na\"ive approach to this setting is to simply
couple all time series under a shared HDP prior. However, this approach
assumes that the state spaces of the multiple Markov switching
processes are \emph{exactly} shared, as are the transitions among
these states. As demonstrated in Section~\ref{secMoCap} as well as
our extensive toy data experiments in Supplement~H
of~\citet{BPARHMMSupp}, such strict sharing can limit the ability
to discover unique dynamic behaviors and reduces predictive performance.

In recent independent work, \citet{saria2010discovering}
developed an alternative model for multiple time series via the
HDP-HMM. Their \emph{time series topic model} (TSTM) describes
coarse-scale temporal behavior using a finite set of ``topics,'' which
are themselves distributions on a common set of autoregressive
dynamical models. Each time series is assumed to exhibit all topics to
some extent, but with unique frequencies and temporal patterns.
Alternatively, the mixed HMM~[\citet{altman2007mixed}] uses
generalized linear models to allow the state transition and emission
distributions of a finite HMM to depend on arbitrary external
covariates. In experiments, this is used to model the differing
temporal dynamics of a small set of known time series classes.

More broadly, the problem we address here has received little previous
attention, perhaps due to the difficulty of treating combinatorial
relationships with parametric models. There are a wide variety of
models which capture correlations among multiple aligned, interacting
univariate time series, for example, using Gaussian state space
models~[\citet{aoki1991state}].
Other approaches cluster time series using a parametric mixture
model~[\citet{alon2003discovering}], or a Dirichlet process
mixture~[\citet{qi07}], and model the dynamics within each
cluster via independent finite HMMs.

Dynamic Bayesian networks~[\citet{murphy2002dynamic}], such as
the factorial HMM [\citet{ghahramani97}], define a structured
representation for the latent states underlying a single time series.
Factorial models are widely used in applied time series
analysis~[\citet{lehrach2009segmenting,duh2005jointly}].
The infinite factorial HMM~[\citet{VanGael082}] uses the IBP to
model a single time series via an infinite set of latent features, each
evolving according to independent Markovian dynamics. Our work instead
focuses on discovering behaviors shared across \emph{multiple} time series.

Other approaches do not explicitly model latent temporal dynamics and
instead aim to align time series with consistent global
structure~[\citet{aach2001aligning}].
Motivated by the problem of detecting temporal anomalies, \citet
{listgarten2007bayesian} describe a hierarchical Bayesian approach to
modeling shared structure among a known set of time series classes.
Independent HMMs are used to encode nonlinear alignments of observed
signal traces to latent reference time series, but their states do not
represent dynamic behaviors and are not shared among time series.

\section{Motion capture experiments}
\label{secMoCap}
The linear dynamical system is a common model for describing simple
human motion~[\citet{Hsu05}], and the switching linear dynamical
system (SLDS) has been successfully applied to the problem of human
motion synthesis, classification, and visual tracking~[\citet
{Pavlovic99,Pavlovic01}]. Other approaches develop nonlinear dynamical
models using Gaussian processes~[\citet{Wang08}] or are based on
a collection of binary latent features~[\citet{Taylor07}].
However, there has been little effort in jointly segmenting and
identifying common dynamic behaviors among a set of \emph{multiple}
motion capture (MoCap) recordings of people performing various tasks.
The ability to accurately label frames of a large set of movies is
useful for tasks such as querying an extensive database without relying
on expensive manual labeling.

The BP-AR-HMM provides a natural way to model complex MoCap data, since
it does not require manually specifying the set of possible behaviors.
In this section, we apply this model to sequences from the well-known
CMU MoCap database~[\citet{CMUmocap}]. Using the smaller
6-sequence data set from Figure~\ref{figGroundTruthMocap6}, we first
justify our proposed MCMC algorithm's benefits over prior methods for
inference, and also show improved performance in segmenting these
sequences relative to alternative parametric models. We then perform an
exploratory analysis of a larger 124-sequence MoCap data set.

\subsection{Data preprocessing and hyperparameter selection}
As described in Section~\ref{secdata}, we examine multivariate time
series generated by 12 MoCap sensors. The CMU data are recorded at a
rate of 120 frames per second, and as a preprocessing step we
block-average and downsample the data using a window size of 12. We
additionally scale each component of the observation vector so that the
empirical variance of the set of first-difference measurements, between
observations at neighboring time steps, is equal to one.

We fix the hyperparameters of the MNIW prior on $\theta_k$ in an
empirical Bayesian fashion using statistics derived from the sample
covariance of the observed data.
These settings are similar to prior work [\citet{BPHMMNIPS12}]
and are detailed in Supplement~J of~\citet
{BPARHMMSupp}. The IBP hyperparameters $\alpha,c$ and the transition
hyperparameters $\gamma,\kappa$ are sampled at every iteration [see
Supplement~G of~\citet{BPARHMMSupp}, which
also discusses hyperprior settings].

%
\subsection{Comparison of BP-AR-HMM sampler methods}
%
%
\begin{figure}

\includegraphics{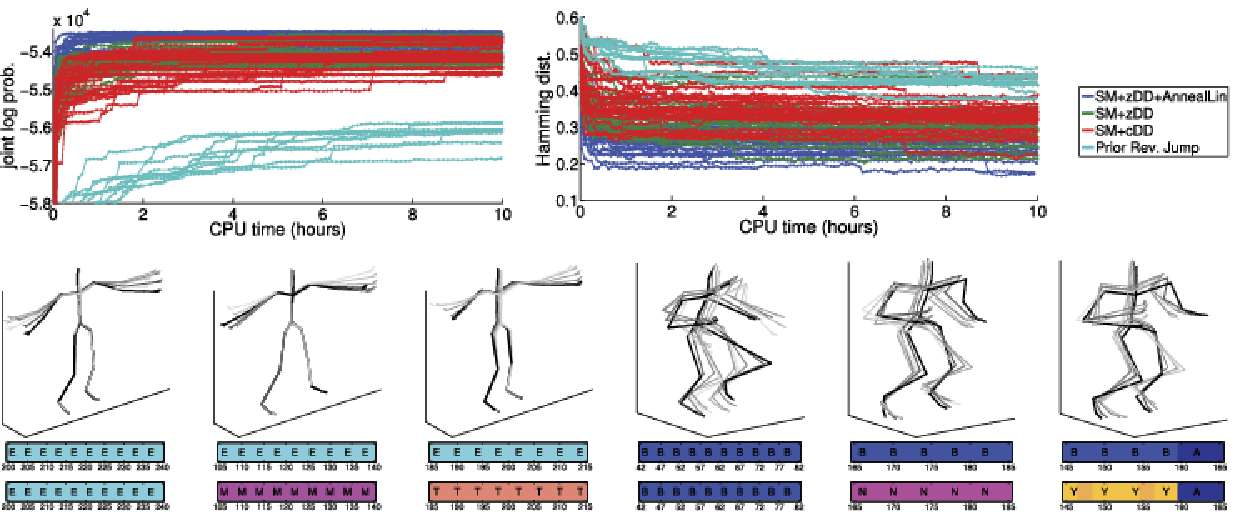}

\caption{Analysis of six MoCap sequences, comparing sampling methods.
Baselines are reversible jump proposals from the prior [\citet
{BPHMM}], and split-merge moves interleaved with data-driven proposals
of continuous parameters (SM${}+{}$cDD) [\citet{BPHMMNIPS12}]. The
proposed sampler interleaves split-merge and data-driven discrete
variable proposals (SM${}+{}$zDD), with and without annealing. \emph{Top
row}: Log-probability and Hamming distance for 25 runs of each method
over 10 hours. \emph{Bottom row}: Estimated state sequence $\mathbf
{z}$ for three fragments from distinct sequences that humans label
``arm circles'' (\emph{left}) or ``jogging'' (\emph{right}).
Each recovered feature is depicted by one unique color and letter.
We compare segmentations induced by the most probable samples from the
annealed SM${}+{}$zDD (\emph{top}) and Prior Rev. Jump (\emph{bottom}) methods.
The latter creates extraneous features.}
\label{figMoCap6}
\end{figure}

Before comparing our BP-AR-HMM to alternative modeling techniques, we
first explore the effectiveness of several possible MCMC methods for
the BP-AR-HMM. As baselines, we implement several previous methods that
use reversible jump procedures and propose moves in the space of
continuous HMM parameters. These include proposals for $\theta_k$ from
the prior [\citet{BPHMM}, ``Prior Rev. Jump''], and split-merge
moves interleaved with data-driven proposals for $\theta_k$
[\citet{BPHMMNIPS12}, ``SM${}+{}$cDD'']. The birth--death moves for
these previous approaches act on the \emph{continuous} HMM parameters,
as detailed in Supplement~E of~\citet
{BPARHMMSupp}. We compare these to our proposed split-merge and
birth--death moves on the discrete assignment variables from
Section~\ref{seczDD} (``SM${}+{}$zDD''). Finally, we consider annealing the
SM${}+{}$zDD moves (Section~\ref{secAnneal}).

We run 25 chains of each method for 10 hours, which allows at least
10,000 iterations for each individual run. All split-merge methods
utilize a parsimonious initialization starting from just a single
feature shared by all sequences. The Prior Rev. Jump algorithm rarely
creates meaningful new features from this simple initialization, so
instead we initialize with five unique features per sequence as
recommended in \citet{BPHMM}. The results are summarized in
Figure~\ref{figMoCap6}. We plot traces of the joint log probability
of data and sampled variables, $p( \mathbf{y}, \mathbf{F}, \mathbf{z},
\alpha, c, \gamma, \kappa)$, versus elapsed wall-clock time. By
collapsing out the continuous HMM parameters $\bolds{\theta},\bolds
{\eta}$, the marginalized form allows direct comparison of
configurations despite possible differences in the number of
instantiated features [see Supplement~C of~\citet
{BPARHMMSupp} for computation details]. We also plot the temporal
evolution of the normalized Hamming distance between the sampled
segmentation $\mathbf{z}$ and the human-provided ground truth
annotation, using the optimal alignment of each ``true'' state to a
sampled feature. Normalized Hamming distance measures the fraction of
time steps where the labels of the ground-truth and estimated
segmentations disagree. To compute the optimal (smallest Hamming
distance) alignment of estimated and true states, we use the Hungarian
algorithm.

With respect to both the log-probability and Hamming distance metrics,
we find that our SM${}+{}$zDD inference algorithm with annealing yields the
best results. Most SM${}+{}$zDD runs using annealing (blue curves) converge
to regions of good segmentations (in terms of Hamming distance) in
under two hours, while no run of the Prior Rev. Jump proposals (teal
curves) comes close after ten hours. This indicates the substantial
benefit of using a data-driven proposal for adding new features
efficiently. We also find that on average our new annealing approach
(blue) improves on the speed of convergence compared to the nonannealed
SM${}+{}$zDD runs (green). This indicates that the Hastings factor penalty
discussed in Section~\ref{secAnneal} is preventing some proposals
from escaping local optima. Our annealing approach offers a practical
workaround to overcome this issue, while still providing valid samples
from the posterior after burn-in.

Our split-merge and data-driven moves are critical for effectively
creating and deleting features to produce quality segmentations. In the
lower half of Figure~\ref{figMoCap6}, we show sampled segmentations
$\mathbf{z}$ for fragments of the time series from distinct sequences
that our human annotation labeled ``arm-circle'' or ``jogging.'' SM${}+{}$zDD
with annealing successfully explains each action with one primary state
reused across all subjects. In contrast, the best Prior Rev. Jump run
(in terms of joint probability) yields a poor segmentation that assigns
multiple unique states for one common action, resulting in lower
probability and much larger Hamming distance. This over-segmentation is
due to the 5-unique-features-per-sequence initialization used for the
Prior Rev. Jump proposal, but we found that a split-merge sampler using
the same initialization could effectively merge the redundant states.
Our merge proposals are thus effective at making global changes to
remove redundant features; such changes are extremely unlikely to occur
via the local moves of \mbox{standard} samplers. Overall, we find that our
data-driven birth--death moves (zDD) allow rapid creation of crucial new
states, while the split-merge moves (SM) enable global improvements to
the overall configuration. 

Even our best segmentations have nearly 20\% normalized Hamming
distance error. To disentangle issues of model mismatch from mixing
rates, we investigated whether the same SM${}+{}$zDD sampler initialized to
the true human segmentations would retain all ground-truth labeled
exercise behaviors after many iterations. (Of course, such checks are
only possible when ground-truth labels are available.) We find that
these runs prefer to delete some true unique features, consistently
replacing ``K: side-bend'' with ``F: twist.'' Manual inspection reveals
that adding missing unique features back into the model actually
decreases the joint probability, meaning the true segmentation is not
quite a global (or even local) mode for the BP-AR-HMM. Furthermore, the
result of these runs after burn-in yield similar joint log-probability
to the best run of our SM${}+{}$zDD sampler initialized to just one feature.
We therefore conclude that our inference procedure is reasonably
effective and that future work should concentrate on improving the
local dynamical model to better capture the properties of unique human
behaviors.


%
\subsection{Comparison to alternative time series models}
%
%
\begin{figure}

\includegraphics{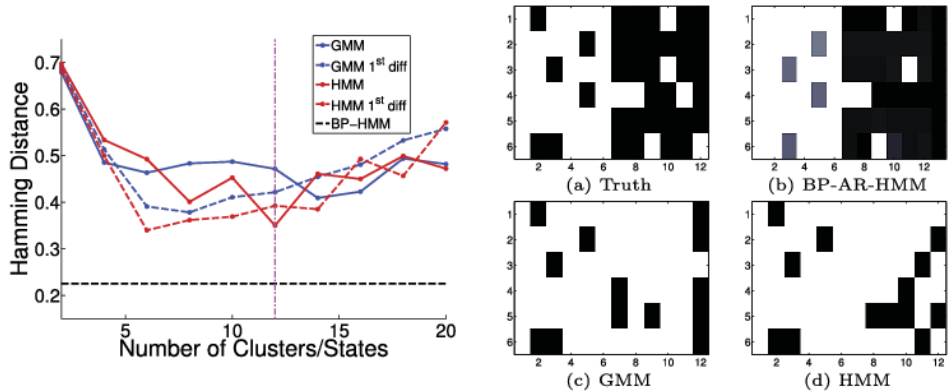}

\caption{Comparison of the BP-AR-HMM analysis using SM${}+{}$zDD inference
(Figure~\protect\ref{figMoCap6}) to parametric HMM and Gaussian
mixture model (GMM) approaches. \emph{Left}: Hamming distance versus
number of GMM clusters/HMM states on raw observations and
first-difference observations, with the BP-AR-HMM segmentation and true
feature count $K=12$ (\textup{magenta}, \textup{vertical dashed}) shown for comparison.
\emph{Right}: Feature matrices associated with \textup{(a)} the human
annotation, \textup{(b)} BP-AR-HMM averaged across MCMC samples, and
maximum-likelihood assignment of the \textup{(c)} GMM and \textup{(d)} HMM using
first-difference observations and 12 states. We set feature $k$ present
in sequence $i$ only if $\mathbf{z}^{(i)}$ is assigned to $k$ for at
least 2\% of its time steps. White indicates a feature being present.}
\label{figCompareOtherModelsMocap}
\end{figure}

We next compare the BP-AR-HMM to alternative models to assess the
suitability of our nonparametric feature-based approach. As
alternatives, we consider the Gaussian mixture model (GMM) method
of~\citet{Barbic04}.\hskip.2pt\footnote{\citet{Barbic04} also
present an approach based on probabilistic principal component analysis
(PCA), but this method focuses primarily on change-point detection
rather than behavior clustering.} We also consider a GMM on
first-difference observations (which behaves like a special case of our
autoregressive model) and an HMM on both first-difference and raw
observations. Note that both the GMM and HMM models are \emph
{parametric}, requiring the number of states to be specified {a
priori}, and that both methods are trained via expectation maximization
(EM) to produce maximum likelihood parameter estimates.

In Figure~\ref{figCompareOtherModelsMocap} we compare all methods'
estimated segmentation accuracy, measuring Hamming distance between the
estimated label sequence $\mathbf{z}$ and human annotation on the six
MoCap sequences. The GMM and HMM results show the most likely of 25
initializations of EM using the HMM Matlab toolbox~[\citet
{MurphyHMMtoolbox}]. Our BP-AR-HMM Hamming distance comes from the best
single MCMC sample (in log probability) among all runs of SM${}+{}$zDD with
annealing in Figure~\ref{figMoCap6}.
The BP-AR-HMM provides more accurate segmentations than the GMM and
HMM, and this remains true regardless of the number of states set for
these parametric alternatives.

The BP-AR-HMM's accuracy is due to better recovery of the sparse
behavior sharing exhibited in the data. This is shown in Figure~\ref
{figCompareOtherModelsMocap}, where we compare estimated binary
feature matrices for all methods. In contrast to the sequence-specific
variability modeled by the BP-AR-HMM, both the GMM and HMM assume that
each sequence uses all possible behaviors, which results in the strong
vertical bands of white in almost all columns. Overall, the BP-AR-HMM
produces superior results due to its flexible feature sharing and
allowance for unique behaviors.

%
%

\subsection{Exploring a large motion capture data set}

\begin{figure}

\includegraphics{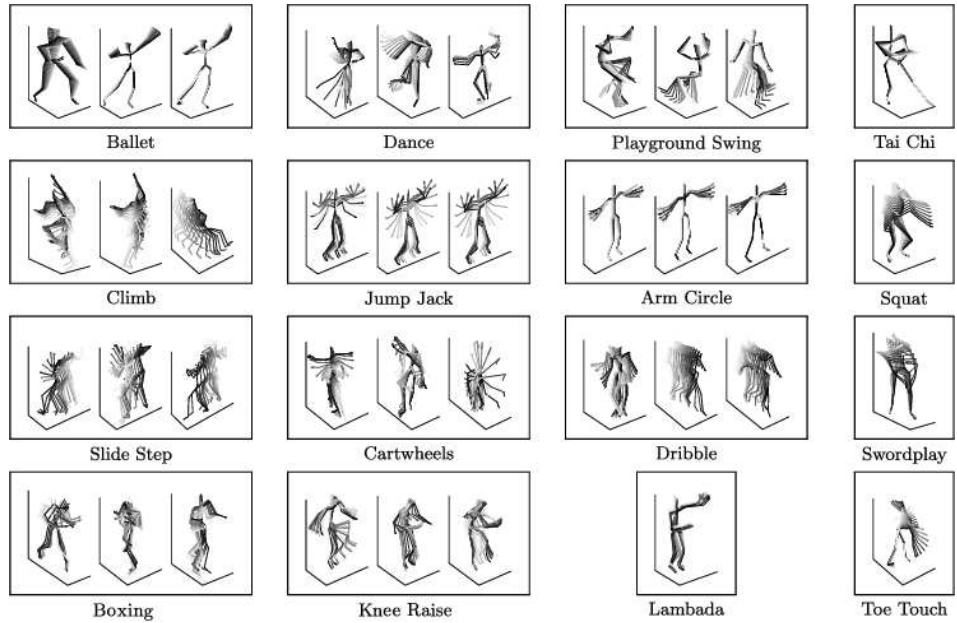}

\caption{Analysis of 124 MoCap sequences by interleaving of
split-merge and data-driven MCMC moves. 16 exemplars of the 33
recovered behaviors are displayed, with text label applied post-hoc to
aid human interpretation. Skeleton trajectories were visualized from
contiguous segments of at least 1 second of data as segmented by the
sampled state sequence $\mathbf{z}^{(i)}$. Boxes group segments from
distinct sequences assigned to the same behavior type.}
\label{figMoCapBIG}
\end{figure}

Finally, we consider a larger motion capture data set of 124 sequences,
all ``Physical Activities \& Sports'' examples from the CMU MoCap data
set (including all sequences in our earlier small data set). The median
length is $T= 95.5$ times steps (minimum 16, maximum~1484).
Human-produced segmentations for ground-truth comparison are not
available for data of this scale. Furthermore, analyzing this data is
computationally infeasible without split-merge and data-driven
birth--death moves. For example, the small data set required a special
``5 unique features per sequence'' initialization to perform well with
Prior Rev. Jump proposals, but using this initialization here would
create over 600 features, requiring a prohibitively long sampling run
to merge related behaviors. In contrast, our full MCMC sampler (SM-zDD
with annealing) completed 2000 iterations in 24 hours. Starting from
just one feature shared by all 124 sequences, our SM${}+{}$zDD moves identify
a diverse set of 33 behaviors in this data set. A set of 16
representative behaviors are shown in Figure~\ref{figMoCapBIG}. The
resulting clusterings of time series segments represent coherent
dynamic behaviors. Note that a full quantitative analysis of the
segmentations produced on this data set is not possible because we lack
manual annotations. Instead, here we simply illustrate that our
improved inference procedure robustly explores the posterior, enabling
this large-scale analysis and producing promising results.

\section{Discussion} \label{secchap5discussion}
We have presented a Bayesian nonparametric framework for discovering
dynamical behaviors common to multiple time series. Our formulation
reposes on the beta process, which provides a prior distribution on
\mbox{overlapping} subsets of binary features. This prior allows both for
commonality and series-specific variability in the use of dynamic
behaviors. We additionally developed an exact sampling algorithm for
the BP-AR-HMM model, as well as novel split-merge moves and data-driven
birth moves which efficiently explore the unbounded feature space. The
utility of our BP-AR-HMM was demonstrated on the task of segmenting a
large set of MoCap sequences. Although we focused on switching VAR
processes, our approach (and sampling algorithms) could also be applied
to other Markov switching processes, such as switching linear dynamical systems.

The idea proposed herein of a feature-based approach to relating
multiple time series is not limited to nonparametric modeling. One
could just as easily employ these ideas within a parametric model that
prespecifies the number of possible dynamic behaviors. We emphasize,
however, that conditioned on the infinite feature vectors of our
BP-AR-HMM, which are guaranteed to be sparse, our model reduces to a
collection of Markov switching processes on a \emph{finite} state
space. The beta process simply allows for flexibility in the overall
number of globally shared behaviors, and computationally we do not rely
on any truncations of this infinite model.

One area of future work is further improving the split-merge proposals.
Despite the clear benefits of these proposals, we found sometimes that
one ``true'' state would be split among several recovered features. The
root of the splitting issue is twofold. One is the issue of mixing,
which the annealing partially addresses, however, the fundamental issue
of maintaining the reversibility of split-merge moves limits the
acceptance rates due to the combinatorial number of configurations. The
second is due to modeling issues. Our model assumes that the dynamic
behavior parameters (i.e., VAR parameters) are identical between time
series and do not change over time. This assumption can be problematic
in grouping related dynamic behaviors and might be addressed via
hierarchical models of behaviors or by ideas similar to those of the
\emph{dependent Dirchlet process}~[\citet
{MacEachern99,Griffin06}] that allows for time-varying parameters.

Overall, the MoCap results appeared to be fairly robust to examples of
only slightly dissimilar behaviors, such as squatting to different
levels or twisting at different rates. However, in cases such as the
running motion where only portions of the body moved in the same way
while others did not, the behaviors can be split (e.g., third jogging
example in Figure~\ref{figMoCap6}). This observation could motivate
\emph{local partition processes} [\citeauthor{Dunson09} (\citeyear{Dunson09,Dunson09b})]
rather than \emph{global partition processes}. That is, our current
model assumes that the grouping of observations into behavior
categories occurs along all components of the observation vector rather
than just a portion (e.g., lower body measurements). Allowing for
greater flexibility in the grouping of observations becomes
increasingly important in high dimensions.



%

%
%
\begin{supplement}
\stitle{Details on prior specification, derivation of MCMC sampler, and further experimental results}
\slink[doi]{10.1214/14-AOAS742SUPP}
\sdatatype{.pdf}
\sfilename{AOAS742\_supp.pdf}
\sdescription{We provide additional background material on our prior
specification, including the beta process, Indian buffet process, and
inverse Wishart and matrix normal distributions. We also detail aspects
of our MCMC sampler, with further information on the birth--death and
split-merge proposals. Finally, we include synthetic data experiments
and details on the settings used for our MoCap experiments.}
\end{supplement}



%

\printaddresses

\begin{thebibliography}{50}

\bibitem[\protect\citeauthoryear{Aach and Church}{2001}]{aach2001aligning}
%
\begin{barticle}[auto:STB|2014/06/18|12:29:53]
\bauthor{\bsnm{Aach},~\bfnm{J.}\binits{J.}} \AND
\bauthor{\bsnm{Church},~\bfnm{G.}\binits{G.}}
(\byear{2001}).
\btitle{Aligning gene expression time series with time warping algorithms}.
\bjournal{Bioinformatics}
\bvolume{17}
\bpages{495--508}.
\end{barticle}
%
\bptok{imsref}%
\endbibitem

\bibitem[\protect\citeauthoryear{Alon et~al.}{2003}]{alon2003discovering}
%
\begin{bincollection}[auto:STB|2014/06/18|12:29:53]
\bauthor{\bsnm{Alon},~\bfnm{J.}\binits{J.}},
\bauthor{\bsnm{Sclaroff},~\bfnm{S.}\binits{S.}},
\bauthor{\bsnm{Kollios},~\bfnm{G.}\binits{G.}} \AND
\bauthor{\bsnm{Pavlovic},~\bfnm{V.}\binits{V.}}
(\byear{2003}).
\btitle{Discovering clusters in motion time-series data}.
In \bbooktitle{Proc. of the IEEE Conference on Computer Vision and
Pattern Recognition (CVPR)}.
\blocation{Madison, WI, USA}.
\end{bincollection}
%
\bptok{imsref}%
\endbibitem

\bibitem[\protect\citeauthoryear{Altman}{2007}]{altman2007mixed}
%
\begin{barticle}[mr]
\bauthor{\bsnm{Altman},~\bfnm{Rachel~MacKay}\binits{R.~M.}}
(\byear{2007}).
\btitle{Mixed hidden {M}arkov models: {A}n extension of the hidden
{M}arkov model to the longitudinal data setting}.
\bjournal{J. Amer. Statist. Assoc.}
\bvolume{102}
\bpages{201--210}.
\bid{doi={10.1198/016214506000001086}, issn={0162-1459}, mr={2345538}}
\end{barticle}
%
\bptok{imsref}%
\endbibitem

\bibitem[\protect\citeauthoryear{Aoki and Havenner}{1991}]{aoki1991state}
%
\begin{barticle}[mr]
\bauthor{\bsnm{Aoki},~\bfnm{Masanao}\binits{M.}} \AND
\bauthor{\bsnm{Havenner},~\bfnm{Arthur}\binits{A.}}
(\byear{1991}).
\btitle{State space modeling of multiple time series}.
\bjournal{Econometric Rev.}
\bvolume{10}
\bpages{1--99}.
\bid{doi={10.1080/07474939108800194}, issn={0747-4938}, mr={1108250}}
\bptnote{check related}%
\end{barticle}
%
\bptok{imsref}%
\endbibitem

\bibitem[\protect\citeauthoryear{Barbi{\v{c}} et~al.}{2004}]{Barbic04}
%
\begin{bincollection}[auto:STB|2014/06/18|12:29:53]
\bauthor{\bsnm{Barbi{\v{c}}},~\bfnm{J.}\binits{J.}},
\bauthor{\bsnm{Safonova},~\bfnm{A.}\binits{A.}},
\bauthor{\bsnm{Pan},~\bfnm{J.-Y.}\binits{J.-Y.}},
\bauthor{\bsnm{Faloutsos},~\bfnm{C.}\binits{C.}},
\bauthor{\bsnm{Hodgins},~\bfnm{J.~K.}\binits{J.~K.}} \AND
\bauthor{\bsnm{Pollard},~\bfnm{N.~S.}\binits{N.~S.}}
(\byear{2004}).
\btitle{Segmenting motion capture data into distinct behaviors}.
In \bbooktitle{Proc. of Graphics Interface}.
\blocation{London, Ontario, Canada}.
\end{bincollection}
%
\bptok{imsref}%
\endbibitem

\bibitem[\protect\citeauthoryear{Beal, Ghahramani and
Rasmussen}{2001}]{Beal02}
%
\begin{bincollection}[auto:STB|2014/06/18|12:29:53]
\bauthor{\bsnm{Beal},~\bfnm{M.}\binits{M.}},
\bauthor{\bsnm{Ghahramani},~\bfnm{Z.}\binits{Z.}} \AND
\bauthor{\bsnm{Rasmussen},~\bfnm{C.}\binits{C.}}
(\byear{2001}).
\btitle{The infinite hidden Markov model}.
In \bbooktitle{Advances in Neural Information Processing Systems
(NIPS) 14}.
\blocation{Vancouver, Canada}.
\end{bincollection}
%
\bptok{imsref}%
\endbibitem

\bibitem[\protect\citeauthoryear{CMU}{2009}]{CMUmocap}
%
\begin{bmisc}[auto:STB|2014/06/18|12:29:53]
\bauthor{CMU}
(\byear{2009}).
\bhowpublished{Carnegie Mellon University graphics lab motion capture database.
Available at \surl{http://mocap.cs.cmu.edu/}.}
\end{bmisc}
%
\bptok{imsref}%
\endbibitem

\bibitem[\protect\citeauthoryear{Dahl}{2005}]{Dahl05}
%
\begin{bmisc}[auto:STB|2014/06/18|12:29:53]
\bauthor{\bsnm{Dahl},~\bfnm{D.~B.}\binits{D.~B.}}
(\byear{2005}).
\bhowpublished{Sequentially-allocated merge-split sampler for
conjugate and nonconjugate dirichlet process mixture models. Technical report,
Texas A\&M Univ., College Station, TX.}
\end{bmisc}
%
\bptok{imsref}%
\endbibitem

\bibitem[\protect\citeauthoryear{Duh}{2005}]{duh2005jointly}
%
\begin{bincollection}[auto:STB|2014/06/18|12:29:53]
\bauthor{\bsnm{Duh},~\bfnm{K.}\binits{K.}}
(\byear{2005}).
\btitle{Jointly labeling multiple sequences: A factorial HMM approach}.
In \bbooktitle{43rd Annual Meeting of the Assoc. for Computational
Linguistics (ACL)}.
\blocation{Ann Arbor, MI}.
\end{bincollection}
%
\bptok{imsref}%
\endbibitem

\bibitem[\protect\citeauthoryear{Dunson}{2009}]{Dunson09}
%
\begin{barticle}[mr]
\bauthor{\bsnm{Dunson},~\bfnm{David~B.}\binits{D.~B.}}
(\byear{2009}).
\btitle{Nonparametric {B}ayes local partition models for random effects}.
\bjournal{Biometrika}
\bvolume{96}
\bpages{249--262}.
\bid{doi={10.1093/biomet/asp021}, issn={0006-3444}, mr={2507141}}
\end{barticle}
%
\bptok{imsref}%
\endbibitem

\bibitem[\protect\citeauthoryear{Dunson}{2010}]{Dunson09b}
%
\begin{barticle}[mr]
\bauthor{\bsnm{Dunson},~\bfnm{David~B.}\binits{D.~B.}}
(\byear{2010}).
\btitle{Multivariate kernel partition process mixtures}.
\bjournal{Statist. Sinica}
\bvolume{20}
\bpages{1395--1422}.
\bid{issn={1017-0405}, mr={2777330}}
\end{barticle}
%
\bptok{imsref}%
\endbibitem

\bibitem[\protect\citeauthoryear{Fox et~al.}{2009}]{BPHMM}
%
\begin{bincollection}[auto:STB|2014/06/18|12:29:53]
\bauthor{\bsnm{Fox},~\bfnm{E.~B.}\binits{E.~B.}},
\bauthor{\bsnm{Sudderth},~\bfnm{E.~B.}\binits{E.~B.}},
\bauthor{\bsnm{Jordan},~\bfnm{M.~I.}\binits{M.~I.}} \AND
\bauthor{\bsnm{Willsky},~\bfnm{A.~S.}\binits{A.~S.}}
(\byear{2009}).
\btitle{Sharing features among dynamical systems with beta processes}.
In \bbooktitle{Advances in Neural Information Processing Systems
(NIPS) 22}.
\blocation{Vancouver, Canada}.
\end{bincollection}
%
\bptok{imsref}%
\endbibitem

\bibitem[\protect\citeauthoryear{Fox et~al.}{2010}]{FoxIEEESPM}
%
\begin{barticle}[auto:STB|2014/06/18|12:29:53]
\bauthor{\bsnm{Fox},~\bfnm{E.~B.}\binits{E.~B.}},
\bauthor{\bsnm{Sudderth},~\bfnm{E.~B.}\binits{E.~B.}},
\bauthor{\bsnm{Jordan},~\bfnm{M.~I.}\binits{M.~I.}} \AND
\bauthor{\bsnm{Willsky},~\bfnm{A.~S.}\binits{A.~S.}}
(\byear{2010}).
\btitle{Bayesian nonparametric methods for learning Markov switching
processes}.
\bjournal{IEEE Signal Process. Mag.}
\bvolume{27}
\bpages{43--54}.
\end{barticle}
%
\bptok{imsref}%
\endbibitem

\bibitem[\protect\citeauthoryear{Fox et~al.}{2011a}]{FoxIEEE11}
%
\begin{barticle}[auto:STB|2014/06/18|12:29:53]
\bauthor{\bsnm{Fox},~\bfnm{E.~B.}\binits{E.~B.}},
\bauthor{\bsnm{Sudderth},~\bfnm{E.~B.}\binits{E.~B.}},
\bauthor{\bsnm{Jordan},~\bfnm{M.~I.}\binits{M.~I.}} \AND
\bauthor{\bsnm{Willsky},~\bfnm{A.~S.}\binits{A.~S.}}
(\byear{2011}a).
\btitle{Bayesian nonparametric inference of switching dynamic linear models}.
\bjournal{IEEE Trans. Signal Process.}
\bvolume{59}
\bpages{1569--1585}.
\end{barticle}
%
\bptok{imsref}%
\endbibitem

\bibitem[\protect\citeauthoryear{Fox et~al.}{2011b}]{FoxAOAS11}
%
\begin{barticle}[mr]
\bauthor{\bsnm{Fox},~\bfnm{Emily~B.}\binits{E.~B.}},
\bauthor{\bsnm{Sudderth},~\bfnm{Erik~B.}\binits{E.~B.}},
\bauthor{\bsnm{Jordan},~\bfnm{Michael~I.}\binits{M.~I.}} \AND
\bauthor{\bsnm{Willsky},~\bfnm{Alan~S.}\binits{A.~S.}}
(\byear{2011}b).
\btitle{A sticky HDP--HMM with application to speaker diarization}.
\bjournal{Ann. Appl. Stat.}
\bvolume{5}
\bpages{1020--1056}.
\bid{doi={10.1214/10-AOAS395}, issn={1932-6157}, mr={2840185}}
\end{barticle}
%
\bptok{imsref}%
\endbibitem

\bibitem[\protect\citeauthoryear{Fox et~al.}{2014}]{BPARHMMSupp}
%
\begin{bmisc}[auto:STB|2014/06/18|12:29:53]
\bauthor{\bsnm{Fox},~\bfnm{E.~B.}\binits{E.~B.}},
\bauthor{\bsnm{Hughes},~\bfnm{M.~C.}\binits{M.~C.}},
\bauthor{\bsnm{Sudderth},~\bfnm{E.~B.}\binits{E.~B.}} \AND
\bauthor{\bsnm{Jordan},~\bfnm{M.~I.}\binits{M.~I.}}
(\byear{2014}).
\bhowpublished{Supplement to ``Joint modeling of multiple time series
via the beta process with application to motion capture segmentation.''
DOI:\doiurl{10.1214/14-AOAS742SUPP}.}
\end{bmisc}
%
\bptok{imsref}%
\endbibitem

\bibitem[\protect\citeauthoryear{Frigessi et~al.}{1993}]{Frigessi93}
%
\begin{barticle}[mr]
\bauthor{\bsnm{Frigessi},~\bfnm{Arnoldo}\binits{A.}},
\bauthor{\bparticle{di} \bsnm{Stefano},~\bfnm{Patrizia}\binits{P.}},
\bauthor{\bsnm{Hwang},~\bfnm{Chii-Ruey}\binits{C.-R.}} \AND
\bauthor{\bsnm{Sheu},~\bfnm{Shuenn~Jyi}\binits{S.~J.}}
(\byear{1993}).
\btitle{Convergence rates of the {G}ibbs sampler, the {M}etropolis
algorithm and other single-site updating dynamics}.
\bjournal{J. Roy. Statist. Soc. Ser. B}
\bvolume{55}
\bpages{205--219}.
\bid{issn={0035-9246}, mr={1210432}}
\end{barticle}
%
\bptok{imsref}%
\endbibitem

\bibitem[\protect\citeauthoryear{Ghahramani, Griffiths and
Sollich}{2006}]{Ghahramani06}
%
\begin{bincollection}[auto]
\bauthor{\bsnm{Ghahramani},~\bfnm{Zoubin}\binits{Z.}},
\bauthor{\bsnm{Griffiths},~\bfnm{Thomas~L.}\binits{T.~L.}} \AND
\bauthor{\bsnm{Sollich},~\bfnm{Peter}\binits{P.}}
(\byear{2006}).
\btitle{Bayesian nonparametric latent feature models}.
In \bbooktitle{Proc. of the Eighth Valencia International Meeting on
Bayesian Statistics (Bayesian Statistics 8)}.
\blocation{Alicante, Spain}.
\end{bincollection}
%
\bptok{imsref}%
\endbibitem

\bibitem[\protect\citeauthoryear{Ghahramani and Jordan}{1997}]{ghahramani97}
%
\begin{barticle}[auto:STB|2014/06/18|12:29:53]
\bauthor{\bsnm{Ghahramani},~\bfnm{Z.}\binits{Z.}} \AND
\bauthor{\bsnm{Jordan},~\bfnm{M.~I.}\binits{M.~I.}}
(\byear{1997}).
\btitle{Factorial hidden Markov models}.
\bjournal{Machine Learning}
\bvolume{29}
\bpages{245--273}.
\end{barticle}
%
\bptok{imsref}%
\endbibitem

\bibitem[\protect\citeauthoryear{Green}{1995}]{Green95}
%
\begin{barticle}[mr]
\bauthor{\bsnm{Green},~\bfnm{Peter~J.}\binits{P.~J.}}
(\byear{1995}).
\btitle{Reversible jump {M}arkov chain {M}onte {C}arlo computation and
{B}ayesian model determination}.
\bjournal{Biometrika}
\bvolume{82}
\bpages{711--732}.
\bid{doi={10.1093/biomet/82.4.711}, issn={0006-3444}, mr={1380810}}
\end{barticle}
%
\bptok{imsref}%
\endbibitem

\bibitem[\protect\citeauthoryear{Griffin and Steel}{2006}]{Griffin06}
%
\begin{barticle}[mr]
\bauthor{\bsnm{Griffin},~\bfnm{J.~E.}\binits{J.~E.}} \AND
\bauthor{\bsnm{Steel},~\bfnm{M.~F.~J.}\binits{M.~F.~J.}}
(\byear{2006}).
\btitle{Order-based dependent {D}irichlet processes}.
\bjournal{J. Amer. Statist. Assoc.}
\bvolume{101}
\bpages{179--194}.
\bid{doi={10.1198/016214505000000727}, issn={0162-1459}, mr={2268037}}
\end{barticle}
%
\bptok{imsref}%
\endbibitem

\bibitem[\protect\citeauthoryear{Hjort}{1990}]{Hjort90}
%
\begin{barticle}[mr]
\bauthor{\bsnm{Hjort},~\bfnm{Nils~Lid}\binits{N.~L.}}
(\byear{1990}).
\btitle{Nonparametric {B}ayes estimators based on beta processes in
models for life history data}.
\bjournal{Ann. Statist.}
\bvolume{18}
\bpages{1259--1294}.
\bid{doi={10.1214/aos/1176347749}, issn={0090-5364}, mr={1062708}}
\end{barticle}
%
\bptok{imsref}%
\endbibitem

\bibitem[\protect\citeauthoryear{Hsu, Pulli and Popovi\'c}{2005}]{Hsu05}
%
\begin{bincollection}[auto:STB|2014/06/18|12:29:53]
\bauthor{\bsnm{Hsu},~\bfnm{E.}\binits{E.}},
\bauthor{\bsnm{Pulli},~\bfnm{K.}\binits{K.}} \AND
\bauthor{\bsnm{Popovi\'c},~\bfnm{J.}\binits{J.}}
(\byear{2005}).
\btitle{Style translation for human motion}.
In \bbooktitle{Proc. of the 32nd International Conference on Computer
Graphics and Interactive Technologies (SIGGRAPH)}.
\blocation{Los Angeles, CA}.
\end{bincollection}
%
\bptok{imsref}%
\endbibitem

\bibitem[\protect\citeauthoryear{Hughes, Fox and
Sudderth}{2012}]{BPHMMNIPS12}
%
\begin{bincollection}[auto:STB|2014/06/18|12:29:53]
\bauthor{\bsnm{Hughes},~\bfnm{M.}\binits{M.}},
\bauthor{\bsnm{Fox},~\bfnm{E.~B.}\binits{E.~B.}} \AND
\bauthor{\bsnm{Sudderth},~\bfnm{E.~B.}\binits{E.~B.}}
(\byear{2012}).
\btitle{Effective split merge Monte Carlo methods for nonparametric
models of sequential data}.
In \bbooktitle{Advances in Neural Information Processing Systems
(NIPS) 25}.
\bnote{Lake Tahoe, NV, USA}.
\end{bincollection}
%
\bptok{imsref}%
\endbibitem

\bibitem[\protect\citeauthoryear{Jain and Neal}{2004}]{JainNeal04}
%
\begin{barticle}[mr]
\bauthor{\bsnm{Jain},~\bfnm{Sonia}\binits{S.}} \AND
\bauthor{\bsnm{Neal},~\bfnm{Radford~M.}\binits{R.~M.}}
(\byear{2004}).
\btitle{A split-merge {M}arkov chain {M}onte {C}arlo procedure for the
{D}irichlet process mixture model}.
\bjournal{J. Comput. Graph. Statist.}
\bvolume{13}
\bpages{158--182}.
\bid{doi={10.1198/1061860043001}, issn={1061-8600}, mr={2044876}}
\end{barticle}
%
\bptok{imsref}%
\endbibitem

\bibitem[\protect\citeauthoryear{Jain and Neal}{2007}]{JainNeal07}
%
\begin{barticle}[mr]
\bauthor{\bsnm{Jain},~\bfnm{Sonia}\binits{S.}} \AND
\bauthor{\bsnm{Neal},~\bfnm{Radford~M.}\binits{R.~M.}}
(\byear{2007}).
\btitle{Splitting and merging components of a nonconjugate {D}irichlet
process mixture model}.
\bjournal{Bayesian Anal.}
\bvolume{2}
\bpages{445--472}.
\bid{doi={10.1214/07-BA219}, issn={1931-6690}, mr={2342168}}
\bptnote{check related}%
\end{barticle}
%
\bptok{imsref}%
\endbibitem

\bibitem[\protect\citeauthoryear{Kingman}{1967}]{Kin1967}
%
\begin{barticle}[mr]
\bauthor{\bsnm{Kingman},~\bfnm{J.~F.~C.}\binits{J.~F.~C.}}
(\byear{1967}).
\btitle{Completely random measures}.
\bjournal{Pacific J. Math.}
\bvolume{21}
\bpages{59--78}.
\bid{issn={0030-8730}, mr={0210185}}
\end{barticle}
%
\bptok{imsref}%
\endbibitem

\bibitem[\protect\citeauthoryear{Kingman}{1993}]{Kingman93}
%
\begin{bbook}[mr]
\bauthor{\bsnm{Kingman},~\bfnm{J.~F.~C.}\binits{J.~F.~C.}}
(\byear{1993}).
\btitle{Poisson Processes}.
\bpublisher{Oxford Univ. Press},
\blocation{New York}.
\bid{mr={1207584}}
\end{bbook}
%
\bptok{imsref}%
\endbibitem

\bibitem[\protect\citeauthoryear{Lehrach and
Husmeier}{2009}]{lehrach2009segmenting}
%
\begin{barticle}[mr]
\bauthor{\bsnm{Lehrach},~\bfnm{Wolfgang~P.}\binits{W.~P.}} \AND
\bauthor{\bsnm{Husmeier},~\bfnm{Dirk}\binits{D.}}
(\byear{2009}).
\btitle{Segmenting bacterial and viral DNA sequence alignments with a
trans-dimensional phylogenetic factorial hidden {M}arkov model}.
\bjournal{J. R. Stat. Soc. Ser. C. Appl. Stat.}
\bvolume{58}
\bpages{307--327}.
\bid{doi={10.1111/j.1467-9876.2008.00648.x}, issn={0035-9254}, mr={2750008}}
\end{barticle}
%
\bptok{imsref}%
\endbibitem

\bibitem[\protect\citeauthoryear{Listgarten
et~al.}{2006}]{listgarten2007bayesian}
%
\begin{bincollection}[auto:STB|2014/06/18|12:29:53]
\bauthor{\bsnm{Listgarten},~\bfnm{J.}\binits{J.}},
\bauthor{\bsnm{Neal},~\bfnm{R.}\binits{R.}},
\bauthor{\bsnm{Roweis},~\bfnm{S.}\binits{S.}},
\bauthor{\bsnm{Puckrin},~\bfnm{R.}\binits{R.}} \AND
\bauthor{\bsnm{Cutler},~\bfnm{S.}\binits{S.}}
(\byear{2006}).
\btitle{Bayesian detection of infrequent differences in sets of time
series with shared structure}.
In \bbooktitle{Advances in Neural Information Processing Systems
(NIPS) 19}.
\blocation{Vancouver, Canada}.
\end{bincollection}
%
\bptok{imsref}%
\endbibitem

\bibitem[\protect\citeauthoryear{Liu}{1996}]{Liu96}
%
\begin{barticle}[mr]
\bauthor{\bsnm{Liu},~\bfnm{Jun~S.}\binits{J.~S.}}
(\byear{1996}).
\btitle{Peskun's theorem and a modified discrete-state {G}ibbs sampler}.
\bjournal{Biometrika}
\bvolume{83}
\bpages{681--682}.
\bid{doi={10.1093/biomet/83.3.681}, issn={0006-3444}, mr={1423883}}
\end{barticle}
%
\bptok{imsref}%
\endbibitem

\bibitem[\protect\citeauthoryear{MacEachern}{1999}]{MacEachern99}
%
\begin{bincollection}[auto:STB|2014/06/18|12:29:53]
\bauthor{\bsnm{MacEachern},~\bfnm{S.~N.}\binits{S.~N.}}
(\byear{1999}).
\btitle{Dependent nonparametric processes}.
In \bbooktitle{ASA Proc. of the Section on Bayesian Statistical Science}.
\bpublisher{Amer. Statist. Assoc.}, \blocation{Alexandria, VA}.
\end{bincollection}
%
\bptok{imsref}%
\endbibitem

\bibitem[\protect\citeauthoryear{Meeds et~al.}{2006}]{Meeds07}
%
\begin{bincollection}[auto:STB|2014/06/18|12:29:53]
\bauthor{\bsnm{Meeds},~\bfnm{E.}\binits{E.}},
\bauthor{\bsnm{Ghahramani},~\bfnm{Z.}\binits{Z.}},
\bauthor{\bsnm{Neal},~\bfnm{R.~M.}\binits{R.~M.}} \AND
\bauthor{\bsnm{Roweis},~\bfnm{S.~T.}\binits{S.~T.}}
(\byear{2006}).
\btitle{Modeling dyadic data with binary latent factors}.
In \bbooktitle{Advances in Neural Information Processing Systems
(NIPS) 19}.
\blocation{Vancouver, Canada}.
\end{bincollection}
%
\bptok{imsref}%
\endbibitem

\bibitem[\protect\citeauthoryear{M{\o}rup, Schmidt and
Hansen}{2011}]{Morup2011}
%
\begin{bincollection}[auto:STB|2014/06/18|12:29:53]
\bauthor{\bsnm{M{\o}rup},~\bfnm{M.}\binits{M.}},
\bauthor{\bsnm{Schmidt},~\bfnm{M.~N.}\binits{M.~N.}} \AND
\bauthor{\bsnm{Hansen},~\bfnm{L.~K.}\binits{L.~K.}}
(\byear{2011}).
\btitle{Infinite multiple membership relational modeling for complex networks}.
In \bbooktitle{IEEE International Workshop on Machine Learning for
Signal Processing}.
\blocation{Beijing, China}.
\end{bincollection}
%
\bptok{imsref}%
\endbibitem

\bibitem[\protect\citeauthoryear{Murphy}{1998}]{MurphyHMMtoolbox}
%
\begin{bmisc}[auto:STB|2014/06/18|12:29:53]
\bauthor{\bsnm{Murphy},~\bfnm{K.~P.}\binits{K.~P.}}
(\byear{1998}).
\bhowpublished{Hidden Markov model (HMM) toolbox for MATLAB. Available
at \surl{http://www.cs.ubc.ca/~murphyk/Software/HMM/hmm.html}.}
\end{bmisc}
%
\bptok{imsref}%
\endbibitem

\bibitem[\protect\citeauthoryear{Murphy}{2002}]{murphy2002dynamic}
%
\begin{bmisc}[auto]
\bauthor{\bsnm{Murphy},~\bfnm{Kevin~Patrick}\binits{K.~P.}}
(\byear{2002}).
\bhowpublished{Dynamic {B}ayesian networks: {R}epresentation,
inference and learning.
Ph.D. thesis, Univ. California, Berkeley.}
\bid{mr={2704368}}
\end{bmisc}
%
\bptok{imsref}%
\endbibitem

\bibitem[\protect\citeauthoryear{Pavlovi\'c, Rehg and
MacCormick}{2000}]{Pavlovic01}
%
\begin{bincollection}[auto:STB|2014/06/18|12:29:53]
\bauthor{\bsnm{Pavlovi\'c},~\bfnm{V.}\binits{V.}},
\bauthor{\bsnm{Rehg},~\bfnm{J.~M.}\binits{J.~M.}} \AND
\bauthor{\bsnm{MacCormick},~\bfnm{J.}\binits{J.}}
(\byear{2000}).
\btitle{Learning switching linear models of human motion}.
In \bbooktitle{Advances in Neural Information Processing Systems
(NIPS) 13}.
\blocation{Vancouver, Canada}.
\end{bincollection}
%
\bptok{imsref}%
\endbibitem

\bibitem[\protect\citeauthoryear{Pavlovi\'c et~al.}{1999}]{Pavlovic99}
%
\begin{bincollection}[auto:STB|2014/06/18|12:29:53]
\bauthor{\bsnm{Pavlovi\'c},~\bfnm{V.}\binits{V.}},
\bauthor{\bsnm{Rehg},~\bfnm{J.~M.}\binits{J.~M.}},
\bauthor{\bsnm{Cham},~\bfnm{T.~J.}\binits{T.~J.}} \AND
\bauthor{\bsnm{Murphy},~\bfnm{K.~P.}\binits{K.~P.}}
(\byear{1999}).
\btitle{A dynamic Bayesian network approach to figure tracking using
learned dynamic models}.
In \bbooktitle{Proc. of the 7th IEEE International Conference on
Computer Vision (ICCV)}.
\blocation{Kerkyra, Greece}.
\end{bincollection}
%
\bptok{imsref}%
\endbibitem

\bibitem[\protect\citeauthoryear{Qi, Paisley and Carin}{2007}]{qi07}
%
\begin{barticle}[mr]
\bauthor{\bsnm{Qi},~\bfnm{Yuting}\binits{Y.}},
\bauthor{\bsnm{Paisley},~\bfnm{John~William}\binits{J.~W.}} \AND
\bauthor{\bsnm{Carin},~\bfnm{Lawrence}\binits{L.}}
(\byear{2007}).
\btitle{Music analysis using hidden {M}arkov mixture models}.
\bjournal{IEEE Trans. Signal Process.}
\bvolume{55}
\bpages{5209--5224}.
\bid{doi={10.1109/TSP.2007.898782}, issn={1053-587X}, mr={2469377}}
\end{barticle}
%
\bptok{imsref}%
\endbibitem

\bibitem[\protect\citeauthoryear{Rabiner}{1989}]{Rabiner89}
%
\begin{barticle}[auto:STB|2014/06/18|12:29:53]
\bauthor{\bsnm{Rabiner},~\bfnm{L.~R.}\binits{L.~R.}}
(\byear{1989}).
\btitle{A tutorial on hidden Markov models and selected applications
in speech recognition}.
\bjournal{Proceedings of the IEEE}
\bvolume{77}
\bpages{257--286}.
\end{barticle}
%
\bptok{imsref}%
\endbibitem

\bibitem[\protect\citeauthoryear{Saria, Koller and
Penn}{2010}]{saria2010discovering}
%
\begin{bmisc}[auto:STB|2014/06/18|12:29:53]
\bauthor{\bsnm{Saria},~\bfnm{S.}\binits{S.}},
\bauthor{\bsnm{Koller},~\bfnm{D.}\binits{D.}} \AND
\bauthor{\bsnm{Penn},~\bfnm{A.}\binits{A.}}
(\byear{2010}).
\bhowpublished{Discovering shared and individual latent structure in
multiple time series.
Available at \arxivurl{arXiv:1008.2028}.}
\end{bmisc}
%
\bptok{imsref}%
\endbibitem

\bibitem[\protect\citeauthoryear{Taylor, Hinton and Roweis}{2006}]{Taylor07}
%
\begin{bincollection}[auto:STB|2014/06/18|12:29:53]
\bauthor{\bsnm{Taylor},~\bfnm{G.~W.}\binits{G.~W.}},
\bauthor{\bsnm{Hinton},~\bfnm{G.~E.}\binits{G.~E.}} \AND
\bauthor{\bsnm{Roweis},~\bfnm{S.~T.}\binits{S.~T.}}
(\byear{2006}).
\btitle{Modeling human motion using binary latent variables}.
In \bbooktitle{Advances in Neural Information Processing Systems
(NIPS) 19}.
\blocation{Vancouver, Canada}.
\end{bincollection}
%
\bptok{imsref}%
\endbibitem

\bibitem[\protect\citeauthoryear{Teh et~al.}{2006}]{Teh06}
%
\begin{barticle}[mr]
\bauthor{\bsnm{Teh},~\bfnm{Yee~Whye}\binits{Y.~W.}},
\bauthor{\bsnm{Jordan},~\bfnm{Michael~I.}\binits{M.~I.}},
\bauthor{\bsnm{Beal},~\bfnm{Matthew~J.}\binits{M.~J.}} \AND
\bauthor{\bsnm{Blei},~\bfnm{David~M.}\binits{D.~M.}}
(\byear{2006}).
\btitle{Hierarchical {D}irichlet processes}.
\bjournal{J. Amer. Statist. Assoc.}
\bvolume{101}
\bpages{1566--1581}.
\bid{doi={10.1198/016214506000000302}, issn={0162-1459}, mr={2279480}}
\end{barticle}
%
\bptok{imsref}%
\endbibitem

\bibitem[\protect\citeauthoryear{Thibaux and Jordan}{2007}]{Thibaux07}
%
\begin{bincollection}[auto:STB|2014/06/18|12:29:53]
\bauthor{\bsnm{Thibaux},~\bfnm{R.}\binits{R.}} \AND
\bauthor{\bsnm{Jordan},~\bfnm{M.~I.}\binits{M.~I.}}
(\byear{2007}).
\btitle{Hierarchical beta processes and the Indian buffet process}.
In \bbooktitle{Proc. of the Eleventh International Conference on
Artificial Intelligence and Statistics (AISTATS)}.
\blocation{San Juan, Puerto Rico}.
\end{bincollection}
%
\bptok{imsref}%
\endbibitem

\bibitem[\protect\citeauthoryear{Tierney}{1994}]{Tierney}
%
\begin{barticle}[mr]
\bauthor{\bsnm{Tierney},~\bfnm{Luke}\binits{L.}}
(\byear{1994}).
\btitle{Markov chains for exploring posterior distributions}.
\bjournal{Ann. Statist.}
\bvolume{22}
\bpages{1701--1762}.
\bid{doi={10.1214/aos/1176325750}, issn={0090-5364}, mr={1329166}}
\end{barticle}
%
\bptok{imsref}%
\endbibitem

\bibitem[\protect\citeauthoryear{Tu and Zhu}{2002}]{tu02}
%
\begin{barticle}[auto:STB|2014/06/18|12:29:53]
\bauthor{\bsnm{Tu},~\bfnm{Z.}\binits{Z.}} \AND
\bauthor{\bsnm{Zhu},~\bfnm{S.~C.}\binits{S.~C.}}
(\byear{2002}).
\btitle{Image segmentation by data-driven Markov chain Monte Carlo}.
\bjournal{IEEE Trans. Pattern Anal. Mach. Intell.}
\bvolume{24}
\bpages{657--673}.
\end{barticle}
%
\bptok{imsref}%
\endbibitem

\bibitem[\protect\citeauthoryear{Van~Gael, Teh and
Ghahramani}{2009}]{VanGael082}
%
\begin{bincollection}[auto:STB|2014/06/18|12:29:53]
\bauthor{\bsnm{Van Gael},~\bfnm{J.}\binits{J.}},
\bauthor{\bsnm{Teh},~\bfnm{Y.~W.}\binits{Y.~W.}} \AND
\bauthor{\bsnm{Ghahramani},~\bfnm{Z.}\binits{Z.}}
(\byear{2009}).
\btitle{The infinite factorial hidden Markov model}.
In \bbooktitle{Advances in Neural Information Processing Systems
(NIPS) 21}.
\blocation{Vancouver, Canada}.
\end{bincollection}
%
\bptok{imsref}%
\endbibitem

\bibitem[\protect\citeauthoryear{Wang and Blei}{2012}]{HDPsplitmerge}
%
\begin{bmisc}[auto:STB|2014/06/18|12:29:53]
\bauthor{\bsnm{Wang},~\bfnm{C.}\binits{C.}} \AND
\bauthor{\bsnm{Blei},~\bfnm{D.}\binits{D.}}
(\byear{2012}).
\bhowpublished{A split-merge MCMC algorithm for the hierarchical
Dirichlet process.
Available at \arxivurl{arXiv:1201.1657}.}
\end{bmisc}
%
\bptok{imsref}%
\endbibitem

\bibitem[\protect\citeauthoryear{Wang, Fleet and Hertzmann}{2008}]{Wang08}
%
\begin{barticle}[auto:STB|2014/06/18|12:29:53]
\bauthor{\bsnm{Wang},~\bfnm{J.~M.}\binits{J.~M.}},
\bauthor{\bsnm{Fleet},~\bfnm{D.~J.}\binits{D.~J.}} \AND
\bauthor{\bsnm{Hertzmann},~\bfnm{A.}\binits{A.}}
(\byear{2008}).
\btitle{Gaussian process dynamical models for human motion}.
\bjournal{IEEE Trans. Pattern Anal. Mach. Intell.}
\bvolume{30}
\bpages{283--298}.
\end{barticle}
%
\bptok{imsref}%
\endbibitem

\bibitem[\protect\citeauthoryear{West and Harrison}{1997}]{West}
%
\begin{bbook}[mr]
\bauthor{\bsnm{West},~\bfnm{Mike}\binits{M.}} \AND
\bauthor{\bsnm{Harrison},~\bfnm{Jeff}\binits{J.}}
(\byear{1997}).
\btitle{Bayesian Forecasting and Dynamic Models},
\bedition{2nd} ed.
\bpublisher{Springer},
\blocation{New York}.
\bid{mr={1482232}}
\end{bbook}
%
\bptok{imsref}%
\endbibitem
\end{thebibliography}
\end{document}